\documentclass[a4paper,english]{article}

\usepackage{tikz}

\usepackage{amssymb,amsthm, amsmath,amsfonts}
\usepackage[all]{xy}
\usepackage{graphicx}
\usepackage{subfigure}
\usepackage{microtype}
\usepackage{booktabs}
\usepackage{enumerate}
\usepackage{pgf}
\usepackage[latin1]{inputenc}
\usepackage{verbatim}
\usepackage{authblk}

\xyoption{arc}
\usetikzlibrary{arrows,shapes,calc,positioning}

  \tikzstyle{resident}=[circle,fill=blue,draw=none,text=white]
  \tikzstyle{mutant}=[resident,fill=red]
  \tikzstyle{reproductor}=[circle,very thick, fill=red,draw=green!60!black,text=white]
  \tikzstyle{dead}=[circle,very thick,fill=blue,draw=black,text=white]
  \definecolor{zzzzzz}{rgb}{0.4,0.4,0.4}
  \definecolor{ffttqq}{rgb}{1,0.2,0}

\allowdisplaybreaks[3]

\let\rs\setminus

\newcommand{\Pp}{\mathbb{P}}

\newcommand{\G}{\mathcal{G}}
\newcommand{\cals}{\mathcal{S}}
\newcommand{\calp}{\mathcal{P}}
\newcommand{\pN}{{\scriptscriptstyle N}}
\newcommand{\uar}{^{\scriptscriptstyle\uparrow}}
\newcommand{\dar}{^{\scriptscriptstyle\downarrow}}
\newcommand{\rar}{^{\scriptscriptstyle\rightarrow}}
\newcommand{\lar}{^{\scriptscriptstyle\leftarrow}}
\newcommand{\invi}[1]{\phantom{#1}}
\DeclareMathOperator{\Aut}{Aut}
\newcommand{\num}[1]{\lvert #1 \rvert}

\newtheorem{theorem}{Theorem}[section]

  \newtheorem{proposition}[theorem]{Proposition}

\theoremstyle{definition}

\newtheoremstyle{named}{}{}{\itshape}{}{\bfseries}{.}{.5em}{\thmnote{#3's }#1}
\theoremstyle{named}

\begin{document}

\date{}

  \setcounter{Maxaffil}{3}
  
    \title{Fast and asymptotic computation of the fixation probability for Moran processes on graphs}       
    \renewcommand\footnotemark{}
    
    \author[a]{Fernando Alcalde Cuesta}
     \author[a,b]{Pablo Gonz\'alez Sequeiros}    
        \author[a,c,d]{\'Alvaro Lozano Rojo}

\AtEndDocument{\bigskip 
\noindent
\textit{E-mail addresses:} \par 
 \addvspace{\medskipamount}
 
{\footnotesize
\noindent
Fernando Alcalde Cuesta:~\texttt{fernando.alcaldecuesta@gmail.com} \par   \addvspace{\smallskipamount}

\noindent
Pablo Gonz\'alez Sequeiros:~\texttt{pablo.gonzalez.sequeiros@usc.es} \par  \addvspace{\smallskipamount}

\noindent
\'Alvaro Lozano Rojo:~\texttt{alvarolozano@unizar.es}  \par  
}}

 \affil[a]{\small ~GeoDynApp - ECSING Group (Spain)}
 \affil[b]{\small ~Departamento de Did\'actica das Ciencias Experimentais, Facultade de Formaci\'on do Profesorado, 
Universidade de Santiago de Compostela, Avda. Ram\'on Ferreiro 10, E-27002 Lugo (Spain)
      }
 \affil[c]{\small ~Centro Universitario de la Defensa, Academia General Militar, Ctra. Huesca s/n. \newline E-50090 Zaragoza (Spain)
      }
\affil[d]{\small ~Instituto Universitario de Matem\'aticas y Aplicaciones, Universidad de Zaragoza (Spain)
      }	

 \maketitle 
 
\begin{abstract}
  Evolutionary dynamics has been classically studied for homogeneous 
      populations, but now there is a growing interest in the non-homogenous  
      case. One of the most important models has been proposed in~\cite{LHN}, 
      adapting to a weighted directed graph the process described in~\cite{M}. 
      The Markov chain associated with the graph can be modified by erasing all 
      non-trivial loops in its state space, obtaining the  so-called Embedded 
      Markov chain (EMC). The fixation probability remains unchanged, but the 
      expected time to absorption (fixation or extinction) is reduced. In this 
      paper, we shall use this idea to compute asymptotically the average 
      fixation probability for complete bipartite graphs $K_{n,m}$. To this 
      end, we firstly review some recent results on evolutionary dynamics on 
      graphs trying to clarify some points. We also revisit the `Star Theorem' 
      proved in \cite{LHN} for the star graphs $K_{1,m}$. Theoretically, EMC 
      techniques allow fast computation of the fixation probability, but in 
      practice this is not always true. Thus, in the last part of the paper, we 
      compare this algorithm with the standard Monte Carlo method for some kind 
      of complex networks.\medskip 

 {\footnotesize

}
\end{abstract}


\section{Introduction and motivation}
\label{introduccion}

Population genetics studies the genetic composition of biological populations, 
and the changes in this composition that result from the action of four 
different processes: \emph{natural selection}, \emph{random drift}, 
\emph{mutation} and \emph{migration}. The \emph{modern evolutionary synthesis} 
combines Darwin's thesis on natural selection and Mendel's theory of 
inheritance. According to this synthesis, the central object of study in 
evolutionary dynamics is the frequency distribution of the alternative forms 
(\emph{allele}) that a hereditary unit (\emph{gene}) can take in a population 
evolving under these forces. 

Many mathematical models have been proposed to understand evolutionary process. 
Introduced in~\cite{M}, the \emph{Moran model}
describes the  change of gene frequency by random drift on a population of finite fixed size. 
This model has many variants, but we assume for simplicity that involved 
organisms are haploids with only two possible alleles $a$ and $A$ for a given 
locus. Suppose there is a single individual with  a copy of the allele $A$. At 
each unit of time, one individual is chosen at random for reproduction and its 
clonal offspring replaces another individual chosen at random to die. 
To model natural selection, individuals with the advantageous allele $A$ are 
assumed to have relative fitness $r>1$ as compared with those with allele $a$ 
of fitness $1$.

Evolutionary dynamics has been classically studied for homogeneous populations, 
but it is a natural question to ask how non-homogeneous structures affect this 
dynamics. In~\cite{LHN}, a generalisation of the Moran process was introduced 
by arranging the population on a directed graph, see also~\cite{N}, 
\cite{SR2} and \cite{ShakarianBio}. In this model, each vertex represents an 
individual in the population, and the offspring of each individual only replace 
direct successors, i.e. end-points of edges with origin in this vertex. The 
\emph{fitness} of an individual represents again its reproductive rate which 
determines how often offspring takes over its neighbour vertices, although 
these vertices do not have to be replaced in an equiprobable way. The 
evolutionary process is described by the choice of stochastic matrix 
$W = (w_{ij})$ where $w_{ij}$ denotes the probability that individual $i$ 
places its offspring into vertex $j$. 
In fact, further generalisations can be considered assuming 
that the probability above is proportional to the product of a weight $w_{ij}$ 
and the fitness of the individual $i$. In this case, $W$ does not need to be 
stochastic, but non-negative. 
The \emph{fixation probability} of the single individual $i$ is the probability that the progeny of $i$ takes over the whole population.
Several interesting and important results are 
shown in~\cite{LHN}: 
\begin{list}{$\bullet$}{\leftmargin=1em}

  \item Different graph structures support different dynamical behaviours 
    amplifying or suppressing the reproductive advantage of \emph{mutant} 
    individuals (with the advantageous allele $A$) over the \emph{resident} 
    individuals (with the disadvantageous allele $a$). 
  
  \item An evolutionary process on a weighted directed graph $(G,W)$ is 
    \emph{equivalent to a Moran process} (i.e. there is a fixation probability 
    well-defined for any 	individual, which coincides with the fixation 
    probability in a homogeneous population) if and only if $(G,W)$ is 
    \emph{weight-balanced}, i.e.
    for any vertex  $i$ the sum of the weights of entering edges $w_-(i) =  
    \sum_{j=1}^N w_{ji}$ and that of leaving edges 
    $w_+(i) = \sum_{j=1}^N w_{ij}$ are equal. This is called the 
    \emph{Circulation Theorem} in~\cite{LHN} and~\cite{N}. 

\end{list}

\noindent
As in the classical setting, mutant individuals will either become extinct or 
take over the whole population, reaching one of the two \emph{absorption} 
states (\emph{extinction} or \emph{fixation}), when a finite population is 
arranged on an undirected graph or on a \emph{strongly connected} directed 
graph (where two different vertices are always connected by an edge-path).  
Even in the first case, the fixation probability depends usually on the 
starting position of the mutant. The effect of this initial placement on mutant 
spread has been discussed in~\cite{BRS1,BRS2}.
\medskip 

In the present paper, we start by summarising some fundamental ideas and 
results on evolutionary dynamics on graphs. In this context, most work involves 
computing the (average) fixation probability, but doing so in general requires 
solving a system of $2^N$ linear equations. In the example of the 
\emph{star graph} described in~\cite{LHN}, like for other examples described 
in~\cite{BR}, \cite{H} and~\cite{LHN}, a high degree of symmetry reduces the 
size of the linear system to a set of $2N$ equations, which becomes 
asymptotically equivalent to a linear system with $N$ equations. We revisit 
this example that will be useful in addressing the study of complete bipartite 
graphs.  Another research direction has been to use Monte Carlo techniques to 
implement numerical simulations, but often limited to small 
graphs~\cite{BRS1}, small random modification of regular graphs~\cite{RS} or 
graphs evolving under random drift~\cite{SR1}. 

Our aim is to show how to modify the stochastic process associated with a 
weighted directed graph to simplify the 
evolutionary process both analytically and numerically. 
Recall that an evolutionary process on a weighted 
directed graph $(G,W)$ with $N$ vertices is a Markov chain with $2^N$ states representing the vertex sets 
inhabited by mutant individuals and transition matrix $P$ derived from $W$. The non-zero entries of $P$ 
can be used to see the state space as a (weighted) directed graph.
We call \emph{loop-erasing} the loop suppression in this graph $\cals$, 
avoiding to remain in the same state  in two consecutive steps and providing 
the \emph{Embedded Markov chain} (EMC) associated to the process. This 
technique is used here to compute asymptotically the average fixation 
probability for complete bipartite graphs, generalising the Star Theorem 
of~\cite{LHN}, see also~\cite{Banerjee}, \cite{Vallade} and~\cite{TanLu}. 
Expected time to \emph{absorption} (fixation or extinction) of 
this EMC has been studied for circular, complete and star graphs in~\cite{H}. 
Here we compare numerically the expected absorption time of both chains on some 
kinds of complex networks. This method can be combined with other approximation 
methods (like the FPRAS method described in~\cite{D&al} for undirected graphs) 
to obtain a fast approximation scheme. 
\medskip

The paper is organised as follows. In Section~2, we review the Moran 
model for homogeneous and non-homogeneous populations. In  Section~3, we revisit the Star Theorem giving an alternative proof of it.
In Section~4, we briefly explain the machinery of the loop-erasing method and 
we use this idea to describe the asymptotic behaviour of the fixation 
probability on the complete bipartite graphs family. At the end, in Section~5,
we include some numerical experiments to evaluate the performance of the 
Monte Carlo method on both the standard and the loop-erased chains for 
different complex networks. 
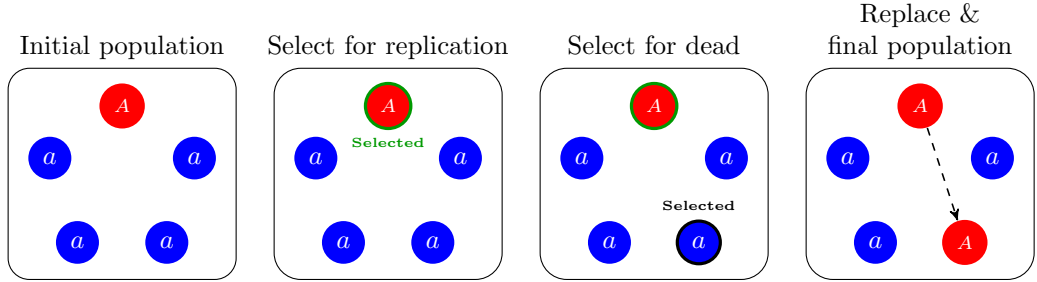
\begin{figure}
  \centering
  \begin{tikzpicture}

    \begin{scope}[xshift=-7cm]
      \node[above] at (0,1.5cm) {Initial population};

      \node[mutant] at (90:1) {$\scriptstyle A$};
      \foreach \a in {1,2,3,4}
        \node[resident] at (90+\a*72:1) {$a$};
        
      \node[draw,rounded corners=8pt,minimum width=3cm,minimum height=2.8cm] 
              at (0,0.1cm) {};
    \end{scope}
    
    \begin{scope}[xshift=-3.5cm]
      \node[above] at (0,1.5cm) {Select for replication};

      \node[reproductor] at (90:1) {$\scriptstyle A$};
      \node[green!60!black, below] at ($(90:1)-(0,.3)$) {\bf\tiny Selected};
      \foreach \a in {1,2,3,4}
        \node[resident] at (90+\a*72:1) {$a$};
      \node[draw,rounded corners=10pt,minimum width=3cm,minimum height=2.8cm] 
              at (0,0.1cm) {};
    \end{scope}

    \node[above] at (0,1.5cm) {Select for dead\vphantom{p}};

    \node[reproductor] (A) at (90:1) {$\scriptstyle A$};
    \foreach \a in {1,2,4}
      \node[resident] at (90+\a*72:1) {$a$};
    \node[dead] at (306:1){$a$};
    
    \node[above] at (0.99,-0.4) at ($(306:1)+(0,.3)$) {\bf\tiny Selected};
    \node[draw,rounded corners=10pt,minimum width=3cm,minimum height=2.8cm] 
              at (0,0.1cm) {};

    \begin{scope}[xshift=3.5cm]
      \node[above,align=center] at (0,1.5cm) {Replace \&\\final population};

      \node[mutant] (A) at (90:1) {$\scriptstyle A$};
      \foreach \a in {1,2,4}
        \node[resident] at (90+\a*72:1) {$a$};
      \node[mutant] (a3) at (306:1) {$\scriptstyle A$};
      \node[draw,rounded corners=10pt,minimum width=3cm,minimum height=2.8cm] 
              at (0,0.1cm) {};
    \draw[-stealth',dashed,semithick] (A) -- (a3);
    \end{scope}

  \end{tikzpicture}

  \caption{Classical Moran process}
      \label{fig:Moran}
\end{figure}

\section{Review of Moran process}
The \emph{Moran process} models random drift and natural selection for finite 
homogeneous populations~\cite{M}. As indicated before, we consider a haploid 
population of $N$ individuals having only two possible alleles $a$ and $A$ for 
a given locus. At the beginning, all individuals have the allele $a$. Then one 
resident individual is chosen at random and replaced by a mutant having the 
neutral or advantageous allele $A$. At successive steps, one randomly chosen 
individual replicates with probability proportional to the 
fitness $r\geq 1$ and its offspring replaces one individual randomly chosen to 
be eliminated, see Figure~\ref{fig:Moran}. Since the future state depends only 
on the present state, the Moran process is a Markov chain $X_n$ with state 
space $\cals= \{0, \dots , N\}$ representing the number of mutant individuals 
with the allele $A$ at the time step $n$. This is a stationary process because 
the probability 
$P_{i,j} =  \Pp[X_{n+1}=j | X_n = i] 
$ to pass from $i$ to $j$ mutant individuals does not depend on the time $n$.
In fact, the number of mutant individuals can change at most by one at each 
step and hence 
the \emph{transition matrix}  $P = (P_{i,j})$ is a tridiagonal matrix
where $P_{i,j} = 0$ if $j \neq i-1, i, i+1$. As $P_{0,0} = P_{\scriptscriptstyle N,N} = 1$, the states $i=0$ and $i=N$ are 
\emph{absorbing}, whereas the other states are \emph{transient}.

The \emph{fixation probability} of $i$ mutant individuals
\[
  \Phi_i = \Phi_i(r) = \Pp[ \exists n \geq 0 : X_n = N | X_0 = i ]
\]
is the solution of the system of linear equations:
\begin{equation}
  \label{eqMoran}
  \begin{aligned}
    \Phi_0 &=  0 \\
    \Phi_i &=  P_{i,i-1}\Phi_{i-1} + P_{i,i}\Phi_i + P_{i,i+1}\Phi_{i+1} \\
    \Phi_\pN &=  1
  \end{aligned}
\end{equation}
where $P_{i,i} = 1 - P_{i,i-1} - P_{i,i+1}$.
In particular, the probability of a single mutant to reach fixation 
$\Phi_1 = \Phi_1(r)$ is usually referred to as the \emph{fixation probability} 
in short.
To solve~\eqref{eqMoran}, we define $y_i = \Phi_i - \Phi_{i-1}$ which verifies $\sum_{i=1}^N y_i = \Phi_\pN - \Phi_0 = 1$. Then, dividing each side of \eqref{eqMoran} by $P_{i,i+1}$, we have $y_{i+1} = \gamma_i y_i$ where $\gamma_i = P_{i,i-1} / P_{i,i+1}$ is the 
\emph{death-birth rate}.
It  follows 
$y_i = \Phi_1 \prod_{j=1}^{i-1} \gamma_j$, and hence the fixation probability is 
\begin{equation}
  \label{fixationMoran}
  \Phi_1 = \frac{1}{1 + \sum_{i=1}^{N-1} \prod_{j=1}^{i} \gamma_j}.
\end{equation}
See \cite{KT}, \cite{Taylor&al} and \cite{Nowak&al}.


If neither of alleles $a$ and $A$ is advantageous reproductively, the {\em random 
drift} phenomenon is modelled by the Moran process with 
fitness $r = 1$, and \eqref{fixationMoran} becomes $\Phi_1=1/N$. 
On the contrary, if mutant individuals with the allele $A$ have 
fitness $r \neq 1$ according to the hypothesis of {\em natural selection}, 
then $\gamma_i = 1/r$ and therefore
\begin{equation}
\label{Moranselection}
  \Phi_1 =  \frac{1}{1 + \sum_{i=1}^{N-1} r^{-i}} = 
                                \frac{1- r^{-1}}{1 - r^{-N}} \geq 1- \frac 1 r .
\end{equation}

\subsection*{Moran process on graphs}

The Moran process for non-homogenous populations represented by graphs was introduced 
in~\cite{LHN}. Like for finite homogenous populations, the first natural 
question is to determine the chance that the offspring of a mutant individual 
having an advantageous allele spreads through the graph reaching any vertex. 
But this chance depends obviously on the initial position of the individual 
(see~\cite{BRS1,BRS2}) and the global graph structure may 
significantly modify the balance between random drift and natural selection 
observed in homogeneous populations as proved in~\cite{LHN}. 

Let $G = (V,E)$ be a directed graph, where $V = \{1, \dots, N\}$ is the set of 
vertices and $E$ is the set of edges. We assume that $G$ is finite, connected 
and simple graph (without loops or multiple edges). Thus $E$ is a 
subset of $\{ (i,j)\in V\times V \mid i\neq j \}$.
An \emph{evolutionary process on $G$} is again
a Markov chain, but each state is now described by a set of vertices 
$S\in\cals = \calp(V)$ inhabited by mutant individuals having a neutral or advantageous 
allele $A$. This reproductive advantage is measured by the fitness $r \geq 1$. 
The transition probabilities of this Markov chain are defined from a 
non-negative matrix $W = (w_{ij})$ satisfying $w_{ij} = 0 \Leftrightarrow \mbox (i,j) \notin E$. More precisely, the transition 
probability between two states $S,S' \in \cals$ is given by
\begin{equation}
  \label{graphtransition} 
  P_{S,S'} = 
    \begin{cases}
      \dfrac{r \sum_{i \in S} w_{ij} }%
            {r \sum_{i \in S} \sum_{j \in V}  w_{ij} + \sum_{i \in  V \rs S}  \sum_{j \in V} w_{ij} }
      & \text{if $S' \rs S = \{j\}$},  \medskip  \\
      \dfrac{\sum_{i \in V \rs S} w_{ij} }%
            {r \sum_{i \in S} \sum_{j \in V}  w_{ij} + \sum_{i \in  V \rs S}  \sum_{j \in V} w_{ij} } 
      & \text{if $S \rs S' = \{j\}$}, \medskip  \\
      \dfrac{r \sum_{i,j \in S} w_{ij} + \sum_{i,j \in V \rs S} w_{ij} }%
            {r \sum_{i \in S} \sum_{j \in V}  w_{ij} + \sum_{i \in  V \rs S}  \sum_{j \in V} w_{ij} }
      & \text{if $S=S'$},  \medskip \\
      0 & \text{otherwise,}
    \end{cases}
\end{equation}
where $r \sum_{i \in S} \sum_{j \in V} w_{ij} + \sum_{i \in V \rs S} \sum_{j \in V} w_{ij}$ is the sum of the reproductive weights of the mutant and  
resident individuals, equal to $r \num{S} + N - \num{S} = N + (r-1)\num{S}$ when the 
matrix $W$ is stochastic. Note that $\cals$ is the vertex set of a directed graph $\G$ where two states $S$ and $S'$ are joined by an edge if and only if $P_{S,S'} \neq 0$.  
Thus, the \emph{Moran process on a 
weighted directed graph $(G,W)$} is the random walk on $\G$
defined by the $2^N \times 2^N$ stochastic matrix $P = (P_{S,S'})$. 
The \emph{fixation probability} of any set $S$ inhabited by mutant individuals 
\[
  \Phi_S = \Phi_S(G,W,r) =  \Pp[ \exists n \geq 0 : X_n = V | X_0 = S ]
\]
is still obtained as the solution of the linear equation 
\begin{equation} \label{ec:fijacion}
P\Phi = \Phi,
\end{equation}
which is analogous to~\eqref{eqMoran} for the classical Moran process. As in this case, $S = \emptyset$ and $S = V$ are absorbing states, but there may be other states of this type, as well as other recurrent states, so the probability that resident or mutant individuals reach fixation can be  strictly less than $1$. However, 
it is well-known (see \cite[Sec.~III.7]{KT:ISM}) that 
\eqref{ec:fijacion} has a unique solution if the only recurrent states are $\emptyset$ and $V$. Thus, the population will still reach one of the two absorbing states: extinction or fixation of mutant individuals. If there are other recurrent states, absorbing or not, 
\eqref{ec:fijacion} will have further solutions if  no other restrictions are 
imposed, see the two-sources digraph below. But the probability of reaching $V$ 
from those states is $0$, so adding these boundary conditions, the uniqueness 
of the fixation probability remain true.

\begin{figure}
\centering
  \begin{tikzpicture}[scale=0.8]
   
    \node[mutant]   (A0) at (90:1.5)   {$\scriptstyle A$};
    \foreach \a in {1,2,3,4}
      \node[resident] (A\a) at (90+72*\a:1.5) {$a$};
  
    \foreach  \a in {0,1,2,3}
      \foreach \b in {1,2,3,4}{
        \ifnum\a<\b
          \draw [stealth'-stealth',semithick] (A\a) -- (A\b);
        \fi
      }
          \path[use as bounding box] 
          ($(current bounding box.north east)+(1,0)$) rectangle 
          ($(current bounding box.south west)-(1,0)$); 

  \end{tikzpicture}
  \caption{Complete graph}
  \label{fig:complete}
  \end{figure}
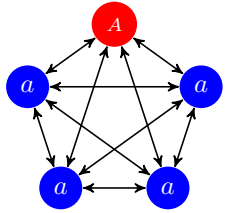

In this context, the fixation probability depends on the starting position of the mutant in the 
graph. This justifies the following definition: 
  for any weighted directed graph $(G,W)$, we call \emph{average fixation probability} 
the average 
$$\Phi_A = \Phi_A(G,W,r) = \frac1N \sum_{i=1}^N \Phi_{\{i\}}.$$
%
\paragraph*{Complete graph} 
Let $K_N$ be the complete graph with vertex set $V = \{1,\dots ,N\}$ and edge 
set $E = \{ (i,j)\in V\times V \mid i\neq j \}$. The classical Moran process is the Moran process on $G = K_N$ defined by the stochastic matrix 
$W = (w_{ij})$ where $w_{ij} = \frac{1}{N-1}$ if $i \neq j$, see 
Figure~\ref{fig:complete}. Since $G$ is \emph{symmetric} (i.e. the automorphism group $\Aut(G)$ acts transitively on $V$ and $E$) and $W$ 
is preserved by the action of $\Aut(G)$, 
$\Phi_{\{i\}} = \Phi_{\{j\}}$ for all $i \neq j$, and then
$\Phi_A = \Phi_{\{i\}}$ for all $i$.

\begin{figure}
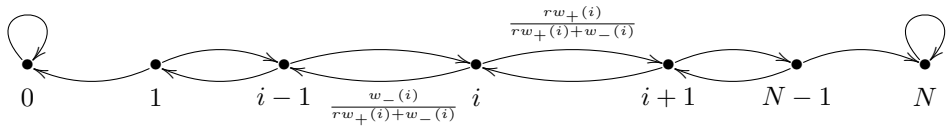

  \centering
  $
    \xy
    0;/r.2pc/: 
    (-70,0)*{\bullet}="A0"; 
    (-70,-5)*{0};
    (-50,0)*{\bullet}="B0"; 
    (-50,-5)*{1};
    (-30,0)*{\bullet}="C0"; 
    (-30,-5)*{i-1};
    (0,0)*{\bullet}="D0"; 
    (0,-5)*{i};
    (30,0)*{\bullet}="E0"; 
    (30,-5)*{i+1};
    (50,0)*{\bullet}="F0"; 
    (50,-5)*{N-1};
    (70,0)*{\bullet}="G0"; 
    (70,-5)*{N};
    {\ar@/^0.5pc/ "B0"; "A0"}; 
    {\ar@/^0.5pc/ "B0"; "C0"}; 
    {\ar@/^0.5pc/ "C0"; "B0"}; 
    {\ar@/^0.5pc/ "C0"; "D0"}; 
    {\ar@/^0.5pc/^{\quad \frac{w_-(i)}{rw_+(i)+w_-(i)}} "D0"; "C0"}; 
    {\ar@/^0.5pc/^{\frac{rw_+(i)}{rw_+(i)+w_-(i)}} "D0"; "E0"}; 
    {\ar@/^0.5pc/ "E0"; "D0"}; 
    {\ar@/^0.5pc/ "E0"; "F0"}; 
    {\ar@/^0.5pc/ "F0"; "E0"}; 
    {\ar@/^0.5pc/ "F0"; "G0"}; 
    {\ar@(ul,ur) "A0"; "A0"}; 
    {\ar@(ul,ur) "G0"; "G0"}; 
    \endxy
$
  \caption{Biased random walk}
  \label{fig:balancedstates}
\end{figure}

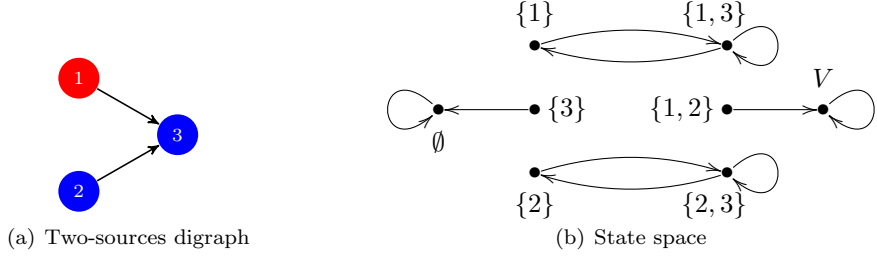
\begin{figure}
  \centering
  \subfigure[Two-sources digraph]{
  \begin{tikzpicture}

    \node[mutant]   (A) at (150:1.5) {$\scriptstyle 1$};
    \node[resident] (B) at (210:1.5) {$\scriptstyle 2$};
    \node[resident] (C) at (0,0)     {$\scriptstyle 3$};
  
    
    \foreach  \from/\to in {A/C, B/C}
      \draw [-stealth', semithick] (\from) -- (\to);

 \path ($(current bounding box.north east)+(0.8,0)$) rectangle 
($(current bounding box.south west)-(0.8,0)$); 

  \end{tikzpicture}
    \label{fig:codirector}
}
\hspace{1cm}
\subfigure[State space]{
\centering
$
    \xy
      0;/r.2pc/: 
      (-35,15)*{\bullet}="A"; 
      (-35,10)*{\emptyset};
      (-20,25)*{\bullet}="B1"; 
      (-20,30)*{\{1\}};
      (-20,15)*{\bullet}="B3"; 
      (-15,15)*{\{3\}};
      (-20,5)*{\bullet}="B2"; 
      (-20,-0)*{\{2\}};
      (10,25)*{\bullet}="C13"; 
      (8,30)*{\{1,3\}};
      (10,15)*{\bullet}="C12"; 
      (3,15)*{\{1,2\}};
      (10,5)*{\bullet}="C23"; 
      (8,0)*{\{2,3\}};
      (25,15)*{\bullet}="D"; 
      (25,20)*{V};
      {\ar@/^0.0pc/ "B3"; "A"}; 
      {\ar@/^0.5pc/ "B1"; "C13"}; 
      {\ar@/^0.5pc/ "C13"; "B1"}; 
      {\ar@/^0.0pc/ "C12"; "D"}; 
      {\ar@/^0.5pc/ "B2"; "C23"}; 
      {\ar@/^0.5pc/ "C23"; "B2"}; 
      {\ar@(lu,dl) "A"; "A"}; 
      {\ar@(ur,dr)  "C13"; "C13"}; 
      {\ar@(ur,dr)  "C23"; "C23"}; 
      {\ar@(ur,dr) "D"; "D"}; 
    \endxy
$
    \label{fig:codirectorstate}
}
\caption{Two-sources digraph and its state space}
\end{figure}

\paragraph*{Weight-balanced graph} 
Assume that $(G,W)$ is 
weight-balanced so that 
 the sum of the weights of entering edges $w_-(i) =  \sum_{j=1}^N w_{ji}$ and that of leaving edges $w_+(i) = \sum_{j=1}^N w_{ij}$ are equal for any vertex 
$i \in V$. According to the Circulation 
Theorem of~\cite{LHN}, 
the number of elements of each state of the Moran process on $(G,W)$ 
\lq performs\rq~a biased random walk on the integer interval $[0,N]$ with 
forward bias $r >1$ and absorbing states $0$ and $N$, see 
Figure~\ref{fig:balancedstates}. Reciprocally, if the Moran process on 
$(G,W)$ reduces to this process,
then $(G,W)$ is weight-balanced.
\paragraph*{Two-sources digraph}
Let $G$ be a directed graph consisting of two vertices (labelled $1$ and $2$) 
having leaving degree $1$ and one vertex (labelled $3$) having entering degree 
$2$, see Figure~\ref{fig:codirector}. There are four recurrent states 
$\{1\}, \{2\},\{1,3\}$ and $\{2,3\}$, the average extinction probability is 
equal to $1/3$, and the average fixation probability is equal to $0$. 
Nonetheless, there is another state $\{1,2\}$ having fixation probability equal 
to $1$, see Figure~\ref{fig:codirectorstate}.
%
\section{Star graphs revisited}
\label{sec:star_reloaded}

\begin{figure}
\centering
  \begin{tikzpicture}[scale=0.85]
   
    \node[mutant] (A)  at (90:1.5)  {$\scriptstyle A$};
    \foreach \a/\b in {1/B, 2/C, 3/D, 4/E}{
      \node[resident] (\b) at (90+72*\a:1.5) {$a$};
    }
    
    \node[resident] (O) at (0,0) {$a$};
  
    \foreach \to in {A, B, C, D, E}
      \draw [stealth'-stealth', semithick] (O) -- (\to);
  
  \end{tikzpicture}
\caption{Star graph}
\label{fig:star}
\end{figure}
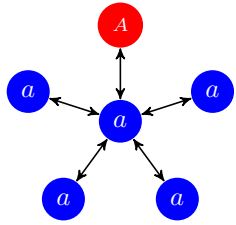

\begin{figure}
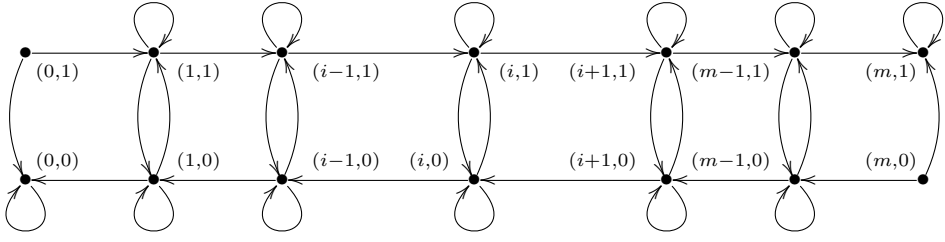

  %
  %
  %
  %
  %
  %
  \centering
  $
    \xy
      0;/r.2pc/: 
      (-70,0)*{\bullet}="A0"; 
      (-65,3)*{\scriptstyle (0,0)};
      (-50,0)*{\bullet}="B0"; 
      (-43,3)*{\scriptstyle (1,0)};
      (-30,0)*{\bullet}="C0"; 
      (-20,3)*{\scriptstyle (i-1,0)};
      (0,0)*{\bullet}="D0"; 
      (-7,3)*{\scriptstyle (i,0)};
      (30,0)*{\bullet}="E0"; 
      (20,3)*{\scriptstyle (i+1,0)};
      (50,0)*{\bullet}="F0"; 
      (40,3)*{\scriptstyle (m-1,0)};
      (70,0)*{\bullet}="G0"; 
      (65,3)*{\scriptstyle (m,0)};
      (-70,20)*{\bullet}="A1"; 
      (-65,17)*{\scriptstyle (0,1)};
      (-50,20)*{\bullet}="B1"; 
      (-43,17)*{\scriptstyle (1,1)};
      (-30,20)*{\bullet}="C1"; 
      (-20,17)*{\scriptstyle (i-1,1)};
      (0,20)*{\bullet}="D1"; 
      (7,17)*{\scriptstyle (i,1)};
      (30,20)*{\bullet}="E1"; 
      (20,17)*{\scriptstyle (i+1,1)};
      (50,20)*{\bullet}="F1"; 
      (40,17)*{\scriptstyle (m-1,1)};
      (70,20)*{\bullet}="G1"; 
      (65,17)*{\scriptstyle (m,1)};
      {\ar@{->} "B0"; "A0"}; 
      {\ar@{->} "C0"; "B0"}; 
      {\ar@{->} "D0"; "C0"}; 
      {\ar@{->} "E0"; "D0"}; 
      {\ar@{->} "F0"; "E0"}; 
      {\ar@{->} "G0"; "F0"}; 
      {\ar@{->} "A1"; "B1"}; 
      {\ar@{->} "B1"; "C1"}; 
      {\ar@{->} "C1"; "D1"}; 
      {\ar@{->} "D1"; "E1"}; 
      {\ar@{->} "E1"; "F1"}; 
      {\ar@{->} "F1"; "G1"}; 
      {\ar@/_0.5pc/ "A1"; "A0"}; 
      {\ar@/_0.5pc/ "G0"; "G1"}; 
      {\ar@/_0.5pc/ "B1"; "B0"}; 
      {\ar@/_0.5pc/ "B0"; "B1"}; 
      {\ar@/_0.5pc/ "C1"; "C0"}; 
      {\ar@/_0.5pc/ "C0"; "C1"}; 
      {\ar@/_0.5pc/@{->} "D1"; "D0"}; 
      {\ar@/_0.5pc/@{->} "D0"; "D1"}; 
      {\ar@/_0.5pc/ "E1"; "E0"}; 
      {\ar@/_0.5pc/ "E0"; "E1"}; 
      {\ar@/_0.5pc/ "F1"; "F0"}; 
      {\ar@/_0.5pc/ "F0"; "F1"}; 
      {\ar@(ul,ur) "B1"; "B1"}; 
      {\ar@(ul,ur) "C1"; "C1"}; 
      {\ar@(ul,ur) "D1"; "D1"}; 
      {\ar@(ul,ur) "E1"; "E1"}; 
      {\ar@(ul,ur) "F1"; "F1"}; 
      {\ar@(ul,ur) "G1"; "G1"}; 
      {\ar@(dr,dl) "A0"; "A0"}; 
      {\ar@(dr,dl) "B0"; "B0"}; 
      {\ar@(dr,dl) "C0"; "C0"}; 
      {\ar@(dr,dl) "D0"; "D0"}; 
      {\ar@(dr,dl) "E0"; "E0"}; 
      {\ar@(dr,dl) "F0"; "F0"}; 
    \endxy
  $
  \caption{State space of a star graph}
  \label{fig:starstates}
\end{figure}

Lieberman et al. showed in \cite{LHN} there are some graph structures, for example \emph{star 
structures}, acting as evolutionary amplifiers favouring advantageous alleles. 
The evolutionary dynamics on stars graphs has been also studied in \cite{BR}. 
We revisit here this example
that is useful to understand the role of symmetry for computing fixation probabilities. 
A \emph{star graph} $G$ 
consists of $N = m+1$ vertices labelled $0,1,\ldots,m$ where only the centre 
$0$ is connected with the peripheral vertices $1,\ldots,m$, see 
Figure~\ref{fig:star}. Since $\Aut(G)$ acts transitively on the peripheral 
vertices, the state space reduces to a set of $2N$ ordered pairs.
The fixation probability of the state $(i,\varepsilon)$ is denoted by
\[
  \Phi_{i,\varepsilon} = 
    \Pp[ \exists n \geq 0 : X_n = (m,1) | X_0 = (i,\varepsilon) ],
\]
where $i$ is the number of peripheral vertices inhabited by mutant individuals 
and $\varepsilon \in \{ 0,1\} $ indicates whether or not there is a mutant
individual at the centre. The evolutionary dynamics of a star structure is 
described by the system of linear equations
\begin{align}
  \Phi_{0,0} &= 0 \nonumber \\
  \Phi_{i,1} &= P_{i,1}^+ \Phi_{i+1,1} +  P_{i,1}^- \Phi_{i,0} \quad + ( 1 - P_{i,1}^+ - P_{i,1}^- ) \Phi_{i,1}  \label{eqstar1} \\
  \Phi_{i,0} &= P_{i,0}^+ \Phi_{i,1} \quad +  P_{i,0}^- \Phi_{i-1,0} + ( 1 - P_{i,0}^+ - P_{i,0}^- ) \Phi_{i,0}  \label{eqstar0} \\
  \Phi_{m,1} &= 1 \nonumber
\end{align}
since 
transitions exist only between state $(i,1)$ (resp. $(i,0)$) and states  
$(i+1,1)$, $(i,0)$ and $(i,1)$ for $i <m$ (resp. $(i-1,0)$, $(i,1)$ and 
$(i,0)$ for $i>0$), see Figure~\ref{fig:starstates}.

The non-trivial entries of $P$ are given by
\begin{align*}
  P_{i,1}^+ &= \Pp[ X_{n+1} = (i+1,1) | X_n = (i,1) ] = \frac{r}{r(i+1)+m-i}\cdot \frac{m-i}{m} \\
  P_{i,1}^- &= \Pp[ X_{n+1} = (i,0)   | X_n = (i,1) ] \invi{+1} = \frac{m-i}{r(i+1)+m-i} \\
  P_{i,0}^+ &= \Pp[ X_{n+1} = (i,1)   | X_n = (i,0) ] \invi{+1} = \frac{ri}{ri+m+1-i} \\
  P_{i,0}^- &= \Pp[ X_{n+1} = (i-1,0) | X_n = (i,0) ] = \frac{1}{ri+m+1-i}\cdot \frac{i}{m}
\end{align*} 
and
\begin{align*}
  1 - P_{i,1}^+ - P_{i,1}^- & = \frac{m+1}{m}\cdot \frac{ri}{r(i+1)+m-i} \\
  1 - P_{i,0}^+ - P_{i,0}^- & = \frac{m+1}{m}\cdot \frac{m-i}{ri+m+1-i} . 
\end{align*}
In particular, we have:
\begin{equation}
  \label{firststep}
  \Phi_{0,1} = \frac{r}{r+m} \Phi_{1,1}
    \qquad\text{and}\qquad 
  \Phi_{1,0} = \frac{rm}{rm +1}\Phi_{1,1}.
\end{equation} 
Thus, the death/birth rates are given by
\[
  \gamma_{i,1} = \frac{P_{i,1}^-}{P_{i,1}^+} = \frac{m}{r}  
    \qquad\text{and}\qquad
  \gamma_{i,0} = \frac{P_{i,0}^-}{P_{i,0}^+} = \frac{1}{rm}.
\]
Like for~\eqref{eqMoran}, the linear equations~\eqref{eqstar1} 
and~\eqref{eqstar0} reduce to 
\begin{align}
  \Phi_{i+1,1} - \Phi_{i,1} &= \gamma_{i,1} (\Phi_{i,1} - \Phi_{i,0}   ) \quad = \frac{m}{r}  (\Phi_{i,1} - \Phi_{i,0}), 
    \label{Eqstar1} \\
  \Phi_{i,1} - \Phi_{i,0}   &= \gamma_{i,0} (\Phi_{i,0} - \Phi_{i-1,0} )  = \frac{1}{rm} ( \Phi_{i,0} - \Phi_{i-1,0} ).
    \label{Eqstar2} 
\end{align}
From~\eqref{Eqstar2}, it is easy to obtain the following identity: 
\begin{equation} 
  \label{Eqstar3}
  \Phi_{i,0} = \sum_{j=1}^i \Bigl(\frac{1}{rm}\Bigr)^{i-j} 
      \Bigl(\frac{rm}{rm+1}\Bigr)^{i-j+1} \Phi_{j,1} , \quad \forall i = 1, \dots, m.
\end{equation}
Now, using~\eqref{Eqstar1} and~\eqref{Eqstar3}, we have the following equation: 
\begin{align*}
  \Phi_{i+1,1} - \Phi_{i,1} &= \frac{m}{r} \biggl[ \Phi_{i,1} - \frac{rm}{rm+1} \Phi_{i,1} - \frac{1}{rm} \cdot \Bigl(\frac{rm}{rm+1}\Bigr)^2 \Phi_{i-1,1} \\
    & \quad - \sum_{j=1}^{i-2} \Bigl(\frac{1}{rm}\Bigr)^{i-j} \Bigl(\frac{rm}{rm+1}\Bigr)^{i-j+1} \Phi_{j,1} \biggr] \\ 
    & = \frac{m}{r(rm+1)} \Phi_{i,1} - \Bigl(\frac{m}{rm+1}\Bigr)^2 \Phi_{i-1,1} \\ 
  & \quad 
    - \sum_{j=1}^{i-2} \frac{m}{r} \Bigl(\frac{1}{rm}\Bigr)^{i-j} \Bigl(\frac{rm}{rm+1}\Bigr)^{i-j+1} \Phi_{j,1}
\end{align*}
where  
\[
  \lim_{m \to +\infty} \sum_{j=1}^{i-2} \frac{m}{r} \Bigl(\frac{1}{rm}\Bigr)^{i-j} \Bigl(\frac{rm}{rm+1}\Bigr)^{i-j+1} \Phi_{j,1} = 0.
\]
Thus, when $m \to +\infty$, the peripheral process 
with fixation probabilities $\Phi_{i,1}$
becomes more and more close to the Moran process 
determined by 
\begin{equation}
  \label{quadratic}
  \Phi_{i+1,1} - \Phi_{i,1}  = \frac{1}{r^2} (\Phi_{i,1} - \Phi_{i-1,1}).
\end{equation} 
According to~\eqref{firststep}, the average fixation probability is
\[
\Phi_A = \frac{1}{m+1} \Phi_{0,1} + \frac{m}{m+1} \Phi_{1,0} = 
  \Bigl( \frac{1}{m+1}\cdot \frac{r}{r+m} + \frac{m}{m+1}\cdot \frac{rm}{rm+1} \Bigr)
  \Phi_{1,1} 
\]
and therefore as $m \to +\infty$, $\Phi_A$ becomes more and more close to the fixation 
probability of the Moran process determined by~\eqref{quadratic} having 
fitness $r^2 > 1$. In short, the star structure is a \emph{quadratic
amplifier of selection}~\cite{LHN} in the sense that the average fixation 
probability of a mutant individual with 
fitness $r > 1$ converges to 
\[
 \Phi_1(r^2) = \frac{1 - r^{-2}}{1- r^{-2m}},
\]
which is  the fixation probability of a mutant with 
fitness $r^2 > 1$ in the Moran process. We will say 
these two 
evolutionary processes are \emph{asymptotically equivalent}. 

\section{Loop-erasing on complete bipartite graphs}

\begin{figure}[t]
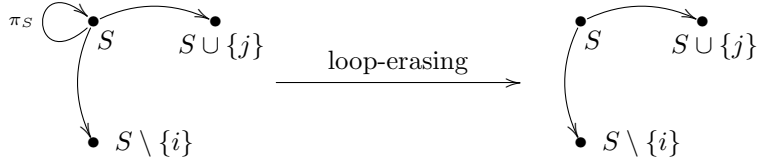

\[
  \xy
    0;/r.19pc/: 
    (-50,0)*{\bullet}="A0"; 
    (-48,-3)*{S};
    (-30,0)*{\bullet}="B0"; 
    (-29,-4)*{S\cup\{j\}};
    (-50,-20)*{\bullet}="C0"; 
    (-40,-20)*{S\setminus\{i\}};
    (-20,-10)*{}="C"; 
    (20,-10)*{}="E"; 
    (30,0)*{\bullet}="E0"; 
    (32,-3)*{S};
    (50,0)*{\bullet}="F0"; 
    (52,-4)*{S\cup\{j\}};
    (30,-20)*{\bullet}="G0"; 
    (40,-20)*{S\setminus\{i\}};
    {\ar@/^0.5pc/ "A0"; "B0"}; 
    {\ar@/_0.5pc/ "A0"; "C0"}; 
    {\ar@/^0.5pc/ "E0"; "F0"}; 
    {\ar@/_0.5pc/ "E0"; "G0"}; 
    {\ar@(dl,ul)^{\pi_S} "A0"; "A0"}; 
    {\ar@/^0.0pc/^{\mbox{loop-erasing}} "C"; "E"}; 
  \endxy
\]
\caption{Loop-erasing method}
\label{fig:looperasing}
\end{figure}

Let us consider a Moran process on a weighted directed graph $(G,W)$.  
This is a random walk on the directed graph $\G$ whose vertex set is $\cals$ and whose transition matrix $P = (P_{S,S'})$ is given by 
\eqref{graphtransition}. Two states $S,S' \in \cals$ are connected by an edge in $\G$ if and only if $P_{S,S'} \neq 0$. 
Let $\hat{\G}$ be the directed graph obtained by suppressing any loop in  $\G$ that connects a non-absorbing state $S$
to itself. For any pair  $S, S' \in \cals$ such that $S$ is non-absorbing,  the transition probability $P_{S,S'}$ is replaced by 
\begin{equation}
  \label{loop-erasing1}
  \hat{P}_{S,S'} = 
  \begin{cases}
    \dfrac{P_{S,S'} }{1-\pi_S} & \text{if $S' \rs S = \{j\}$ or $S \rs S' = \{i\}$},  \medskip  \\
    0 & \text{otherwise,}
  \end{cases}
\end{equation}
where 
\begin{equation}
  \label{loop-erasing2}
  \pi_S = P_{S,S} = 1 - \Bigl( \sum_{j \in V \rs S} P_{S,S\cup\{j\}} 
                                + \sum_{i \in S} P_{S,S \rs \{i\}} \Bigr)
\end{equation}
is the probability of staying one time in the state $S$. Equivalently to \eqref{loop-erasing1}, 
\begin{equation}
  \label{loop-erasing3}
\hat{P}_{S,S'} = \sum_{n \geq 0} \pi_S^n P_{S,S'}
\end{equation}
where the $n$-th power $\pi_S^n$ of $\pi_S$ is the probability of staying $n$ times in the state $S$. We say the random walk on the directed graph $\hat{\G}$ defined by the transition matrix $\hat{P}$ is obtained by 
\emph{loop-erasing} from the Moran process on $(G,W)$, see 
Figure~\ref{fig:looperasing}. This is the \emph{Embedded Markov chain} (EMC) with state space $\cals$ 
obtained by forcing the Moran process on $(G, W)$ to change of state in each step. 
The fixation probability of any set $S$ inhabited by mutant individuals remains 
unchanged 
$\hat{\Phi}_S = \Phi_S$,
because the system of linear equations
\[
\Phi_S = P_{S,S}\Phi_S + \sum_{j \in V \rs S} P_{S,S\cup\{j\}} \Phi_{S\cup\{j\}} 
                       + \sum_{i \in S} P_{S,S \rs \{i\}}  \Phi_{S \rs \{i\}}
\]
can be rewritten as 
\[
\Phi_S = \sum_{j \in V \rs S} \hat{P}_{S,S\cup\{j\}}  \Phi_{S\cup\{j\}} + \sum_{i \in S} \hat{P}_{S,S \rs \{i\}}  \Phi_{S \rs \{i\}}.
\]
The biased random walk described in Figure~\ref{fig:balancedstates} arises by loop-erasing in any process equivalent to the Moran process.

Assuming that $\emptyset$ and $V$ are the only recurrent states in $\G$, we know the population 
will reach one of these two absorbing states, fixation or extinction,
from any other subset $S\subset V$ inhabited by  mutant individuals. Moreover, the transition matrix $P$ admits a box decomposition
\begin{equation}
  \label{block}
  P = \left(\!\begin{array}{c|c|c}
        1 & 0 & 0  \\ \hline
        b & Q & c \\ \hline
        0 & 0 & 1 \\
      \end{array}\!\right).
\end{equation} 
For this type of absorbing Markov chain, the \emph{expected 
absorption time} (i.e. the expected number of steps needed to go from the state 
$S$ to one of the absorbing states $\emptyset$ or $V$) is given by the system 
of linear equations 
\begin{equation}   \label{absorptiontime1}
\tau_S = \sum_{S' \in \cals_T} P_{S,S'} \tau_{S'} + 1
\end{equation} 
where $\cals_T$ is the set of transient states, that is, different from 
$\emptyset$ and $V$. 
Using the box decomposition~\eqref{block}, 
the equation~\eqref{absorptiontime1} reduces to
\begin{equation}
  \label{absorptiontime2}
  \tau = (Id - Q)^{-1}\mathbf{1} = \sum_{n\geq 0} Q^n \mathbf{1},
\end{equation}
where $(Id - Q)^{-1}$ is the fundamental matrix of the Markov chain and 
$\mathbf{1}$ is the vector with all the coordinates equal to $1$. We 
have similar identities for the Markov chain obtained by loop-erasing. Thus, 
using the obvious notation, the new expected absorption time is given by
\begin{equation}
  \label{absorptiontime3}
  \hat{\tau} = \sum_{n\geq 0} \hat{Q}^n \mathbf{1},
\end{equation}
where $(Id- \hat{Q})^{-1} = \sum_{n \geq 0} \hat{Q}^n$. The vector 
$\hat\tau_S$ 
represents the \emph{expected number of state 
transitions until absorption} when the Moran process starts from a set $S$. 
This quantity has been studied in~\cite{H} for circular, complete and star 
graphs. Since transition may not happen at every step, the following 
result is clear:

\begin{proposition}
  \label{thexabtime}
  Let $\tau$ be the expected absorption time for the Moran process on a 
  weighted directed graph $(G,W)$. Let $\hat{\tau}$ be the expected absorption 
  time for the process obtained by applying the loop-erasing method. Then for 
  each transient state $S \in \cals_T$, we have $\hat{\tau}_S \leq \tau_S$.
\end{proposition} 

For unweighted and undirected graphs, D\'{\i}az et al. show in \cite{D&al} that, with high 
probability, the expected absorption time is bounded by a polynomial in $N$ of 
order $3$, $4$ and $6$ when $r <1$, 
$r > 1$ and $r = 1$. They have also 
constructed a fully polynomial randomised approximation scheme for the 
probability of fixation and extinction. The loop-erasing method can be used to 
reduce the expected absorption time making the approximation of the fixation probability faster. We 
explore this path in Section~\ref{sec:experiments}.
\begin{figure}
  \centering
  \subfigure[Complete bipartite graph $K_{2,3}$]{
    \begin{tikzpicture}
      \path[use as bounding box] (-2.5,-2.5) rectangle (2.5,1.25);
   
      \node[mutant]   (A) at ( 0   , 0  ) {$\scriptstyle A$};
      \node[resident] (B) at (-1.5 , 0  ) {$a$};
      \node[resident] (C) at ( 1.5 , 0  ) {$a$};
      \node[resident] (D) at ( 0.75,-1.5) {$a$};
      \node[resident] (E) at (-0.75,-1.5) {$a$};
  
      \foreach \from/\to in {A/D, A/E, B/D, B/E, C/D, C/E}
        \draw [stealth'-stealth', semithick] (\from) -- (\to);
      
    \end{tikzpicture}
    \label{fig:bipartitedif}
  }
  \qquad
  \subfigure[Complete bipartite graph $K_{3,3}$]{
    \begin{tikzpicture}
      \path[use as bounding box] (-2.5,-2) rectangle (2.5,2);
      
      \node[mutant]  (A) at  (90:1.5) {$\scriptstyle A$};
      \foreach \a/\b in {1/B,2/D,3/F,4/E,5/C}{
        \node[resident] (\b) at (90+\a*60:1.5) {$a$};
      }
  
      \foreach  \from/\to in {A/B, A/C, B/D, C/E, D/F, E/F, A/F, B/E,C/D}
        \draw [<->,>=stealth',>=stealth', semithick] (\from) -- (\to);
    
    \end{tikzpicture}
    \label{fig:bipartiteeq}
  }
  \caption{Complete bipartite graphs}
  \label{fig:bipartite}
\end{figure}
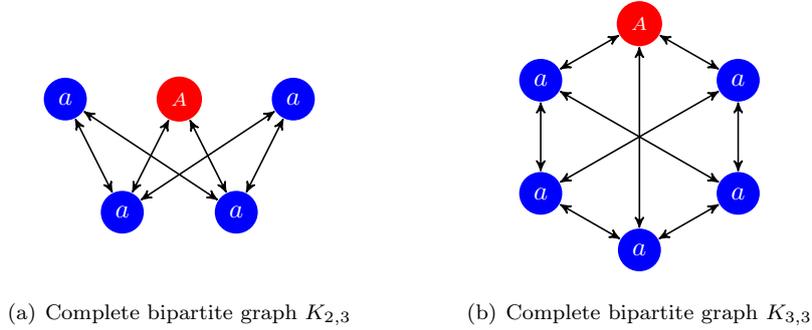

\subsection*{Complete bipartite graph}

Now, we use the loop-erasing method to calculate the asymptotic fixation 
probability of any complete bipartite graph. Recall that a \emph{complete 
bipartite graph} is a graph $K_{m_1,m_2}$ whose vertices can be divided into 
two disjoint sets $V_1 = \{ 1,\ldots,m_1\}$ and $V_2 = \{1,\ldots,m_2\}$ such 
that every edge connects a vertex in  $V_1$ to one in $V_2$. In particular, a 
star graph is a bipartite graph $K_{m,1}$.
The fixation probability for these graphs has been also studied 
in~\cite{Vallade} and~\cite{TanLu}.

\begin{figure}
  \centering
  \subfigure[Average fixation probabilities for 
            $K_{2,10}$, $K_{2,50}$, $K_{2,100}$ and $K_{2,200}$]{
    \includegraphics[width=5cm]{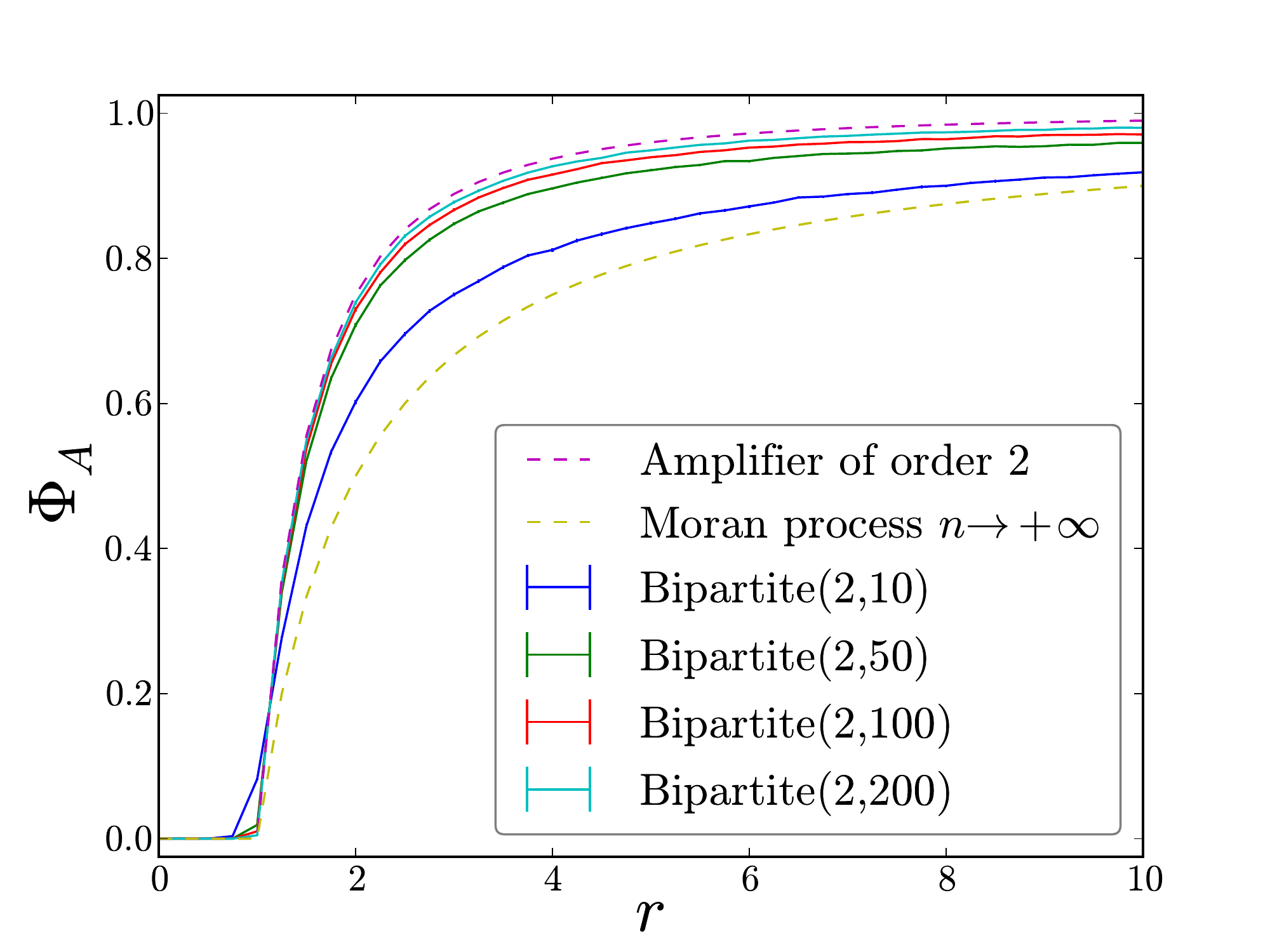}
    \label{fig:Bipartitos2X-MonteCarlo}
  }
  \qquad
  \subfigure[Average fixation probabilities for 
            $K_{10,10}$, $K_{10,50}$, $K_{10,100}$ and $K_{10,200}$]{
    \includegraphics[width=5cm]{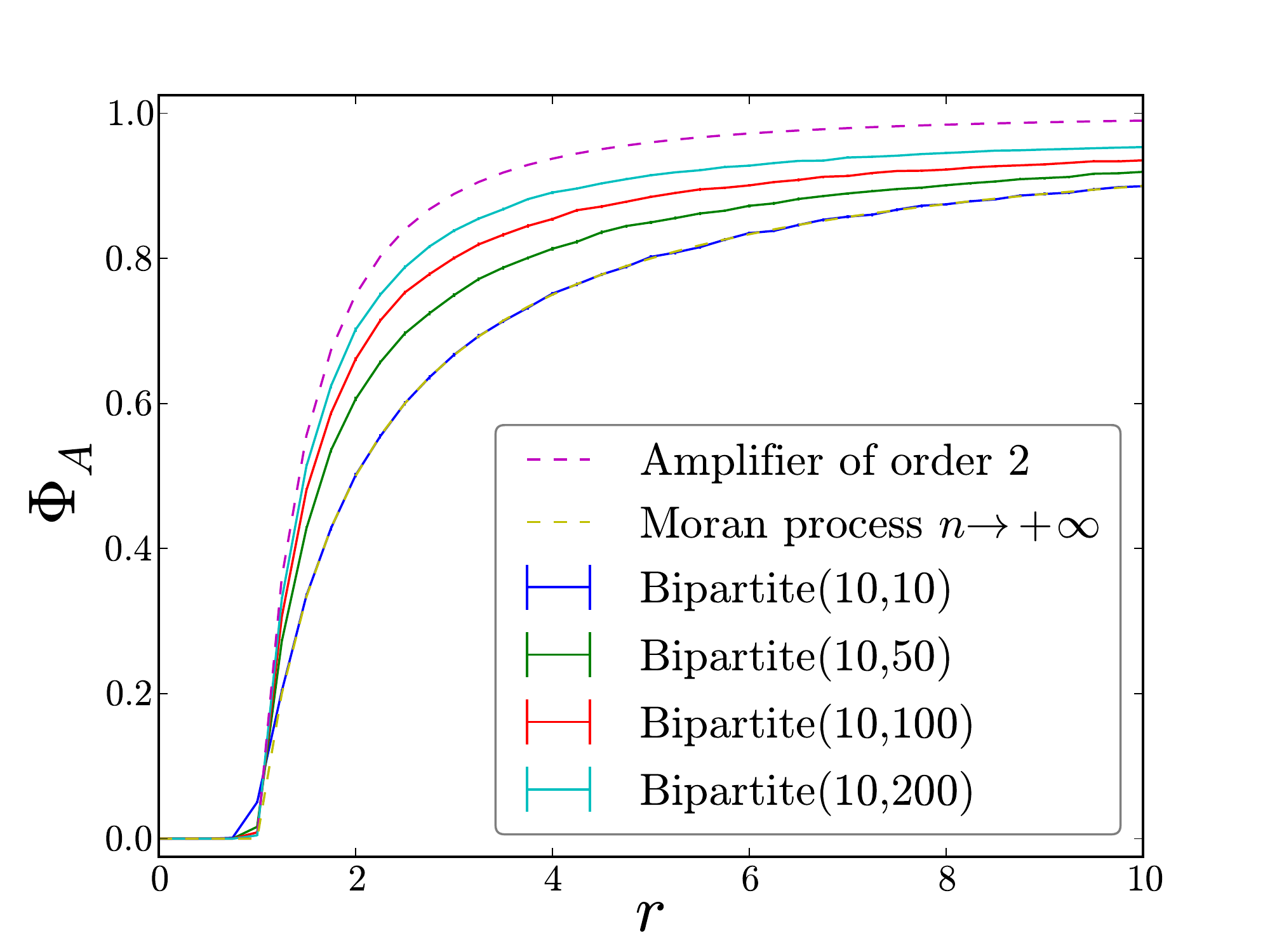}
    \label{fig:Bipartitos10X-MonteCarlo}
  }
  \caption{Average fixation probabilities for Moran processes on some bipartite 
    graphs obtained from Monte Carlo methods. The same figures can be obtained 
    from the loop-erasing method.}
  \label{fig:fixationbipartite}
\end{figure}

According to the Circulation Theorem, as any vertex has the same number of connections, the evolutionary process on the 
complete bipartite graph $K_{m,m}$ is equivalent to the 
 Moran process, so they have the same fixation probability, see 
Figures~\ref{fig:bipartiteeq}~and~\ref{fig:Bipartitos10X-MonteCarlo}. 

\begin{figure}
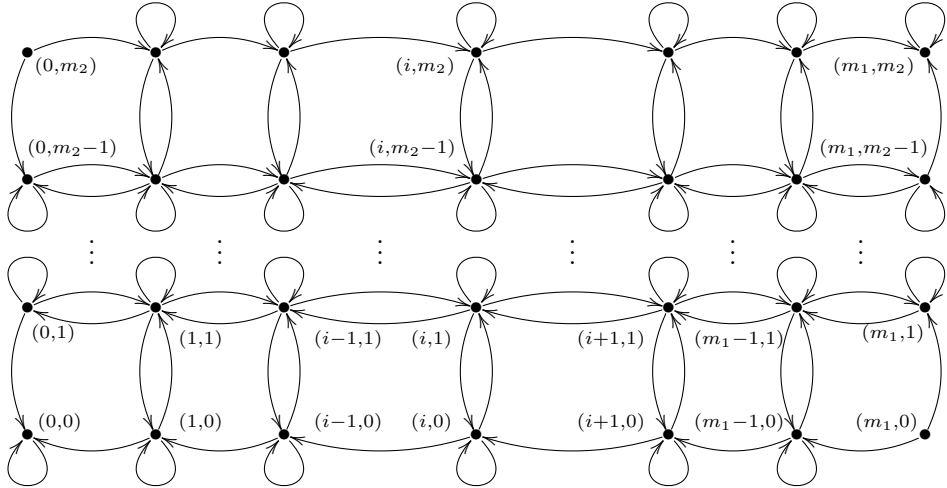

\[
\xy
0;/r.2pc/: 
(-70,0)*{\bullet}="A0"; 
(-65,2)*{\scriptstyle (0,0)};
(-50,0)*{\bullet}="B0"; 
(-43,2)*{\scriptstyle (1,0)};
(-30,0)*{\bullet}="C0"; 
(-20,2)*{\scriptstyle (i-1,0)};
(0,0)*{\bullet}="D0"; 
(-7,2)*{\scriptstyle (i,0)};
(30,0)*{\bullet}="E0"; 
(21,2)*{\scriptstyle (i+1,0)};
(50,0)*{\bullet}="F0"; 
(41,2)*{\scriptstyle (m_1-1,0)};
(70,0)*{\bullet}="G0"; 
(64,2)*{\scriptstyle (m_1,0)};
(-70,20)*{\bullet}="A1"; 
(-66,16)*{\scriptstyle (0,1)};
(-50,20)*{\bullet}="B1"; 
(-43,15)*{\scriptstyle (1,1)};
(-30,20)*{\bullet}="C1"; 
(-20,15)*{\scriptstyle (i-1,1)};
(0,20)*{\bullet}="D1"; 
(-7,15)*{\scriptstyle (i,1)};
(30,20)*{\bullet}="E1"; 
(21,15)*{\scriptstyle (i+1,1)};
(50,20)*{\bullet}="F1"; 
(41,15)*{\scriptstyle (m_1-1,1)};
(70,20)*{\bullet}="G1"; 
(65,16)*{\scriptstyle (m_1,1)};
(-70,40)*{\bullet}="A2"; 
(-63,45)*{\scriptstyle (0,m_2-1)};
(-50,40)*{\bullet}="B2"; 
(-30,40)*{\bullet}="C2"; 
(0,40)*{\bullet}="D2"; 
(-10,45)*{\scriptstyle (i,m_2-1)};
(30,40)*{\bullet}="E2"; 
(50,40)*{\bullet}="F2"; 
(70,40)*{\bullet}="G2"; 
(62,45)*{\scriptstyle (m_1,m_2-1)};
(-70,60)*{\bullet}="A3"; 
(-64,58)*{\scriptstyle (0,m_2)};
(-50,60)*{\bullet}="B3"; 
(-30,60)*{\bullet}="C3"; 
(0,60)*{\bullet}="D3"; 
(-8,58)*{\scriptstyle (i,m_2)};
(30,60)*{\bullet}="E3"; 
(50,60)*{\bullet}="F3"; 
(70,60)*{\bullet}="G3"; 
(62,58)*{\scriptstyle (m_1,m_2)};
(-60,30)*{\vdots}; 
(-40,30)*{\vdots}; 
(-15,30)*{\vdots};
(15,30)*{\vdots};
(40,30)*{\vdots};
(60,30)*{\vdots};
{\ar@/^0.5pc/  "B0"; "A0"}; 
{\ar@/^0.5pc/ "C0"; "B0"}; 
{\ar@/^0.5pc/  "D0"; "C0"}; 
{\ar@/^0.5pc/ "E0"; "D0"}; 
{\ar@/^0.5pc/  "F0"; "E0"}; 
{\ar@/^0.5pc/  "G0"; "F0"}; 
{\ar@/^0.5pc/  "A1"; "B1"}; 
{\ar@/^0.5pc/  "B1"; "C1"}; 
{\ar@/^0.5pc/  "C1"; "D1"}; 
{\ar@/^0.5pc/  "D1"; "E1"}; 
{\ar@/^0.5pc/  "E1"; "F1"}; 
{\ar@/^0.5pc/  "F1"; "G1"}; 
{\ar@/^0.5pc/  "B1"; "A1"}; 
{\ar@/^0.5pc/  "C1"; "B1"}; 
{\ar@/^0.5pc/  "D1"; "C1"}; 
{\ar@/^0.5pc/  "E1"; "D1"}; 
{\ar@/^0.5pc/  "F1"; "E1"}; 
{\ar@/^0.5pc/  "G1"; "F1"}; 
{\ar@/^0.5pc/  "A2"; "B2"}; 
{\ar@/^0.5pc/  "B2"; "C2"}; 
{\ar@/^0.5pc/  "C2"; "D2"}; 
{\ar@/^0.5pc/  "D2"; "E2"}; 
{\ar@/^0.5pc/  "E2"; "F2"}; 
{\ar@/^0.5pc/  "F2"; "G2"}; 
{\ar@/^0.5pc/  "B2"; "A2"}; 
{\ar@/^0.5pc/  "C2"; "B2"}; 
{\ar@/^0.5pc/  "D2"; "C2"}; 
{\ar@/^0.5pc/  "E2"; "D2"}; 
{\ar@/^0.5pc/  "F2"; "E2"}; 
{\ar@/^0.5pc/  "G2"; "F2"}; 
{\ar@/^0.5pc/  "A3"; "B3"}; 
{\ar@/^0.5pc/  "B3"; "C3"}; 
{\ar@/^0.5pc/  "C3"; "D3"}; 
{\ar@/^0.5pc/  "D3"; "E3"}; 
{\ar@/^0.5pc/  "E3"; "F3"}; 
{\ar@/^0.5pc/  "F3"; "G3"}; 
{\ar@/_0.5pc/ "A1"; "A0"}; 
{\ar@/_0.5pc/ "G0"; "G1"}; 
{\ar@/_0.5pc/ "B1"; "B0"}; 
{\ar@/_0.5pc/ "B0"; "B1"}; 
{\ar@/_0.5pc/ "C1"; "C0"}; 
{\ar@/_0.5pc/ "C0"; "C1"}; 
{\ar@/_0.5pc/@{->} "D1"; "D0"}; 
{\ar@/_0.5pc/@{->} "D0"; "D1"}; 
{\ar@/_0.5pc/ "E1"; "E0"}; 
{\ar@/_0.5pc/ "E0"; "E1"}; 
{\ar@/_0.5pc/ "F1"; "F0"}; 
{\ar@/_0.5pc/ "F0"; "F1"}; 
{\ar@/_0.5pc/ "A3"; "A2"}; 
{\ar@/_0.5pc/ "G2"; "G3"}; 
{\ar@/_0.5pc/ "B3"; "B2"}; 
{\ar@/_0.5pc/ "B2"; "B3"}; 
{\ar@/_0.5pc/ "C3"; "C2"}; 
{\ar@/_0.5pc/ "C2"; "C3"}; 
{\ar@/_0.5pc/@{->} "D3"; "D2"}; 
{\ar@/_0.5pc/@{->} "D2"; "D3"}; 
{\ar@/_0.5pc/ "E3"; "E2"}; 
{\ar@/_0.5pc/ "E2"; "E3"}; 
{\ar@/_0.5pc/ "F3"; "F2"}; 
{\ar@/_0.5pc/ "F2"; "F3"}; 
{\ar@(ul,ur) "A1"; "A1"}; 
{\ar@(ul,ur) "B1"; "B1"}; 
{\ar@(ul,ur) "C1"; "C1"}; 
{\ar@(ul,ur) "D1"; "D1"}; 
{\ar@(ul,ur) "E1"; "E1"}; 
{\ar@(ul,ur) "F1"; "F1"}; 
{\ar@(ul,ur) "G1"; "G1"}; 
{\ar@(dr,dl) "A0"; "A0"}; 
{\ar@(dr,dl) "B0"; "B0"}; 
{\ar@(dr,dl) "C0"; "C0"}; 
{\ar@(dr,dl) "D0"; "D0"}; 
{\ar@(dr,dl) "E0"; "E0"}; 
{\ar@(dr,dl) "F0"; "F0"}; 
{\ar@(ul,ur) "B3"; "B3"}; 
{\ar@(ul,ur) "C3"; "C3"}; 
{\ar@(ul,ur) "D3"; "D3"}; 
{\ar@(ul,ur) "E3"; "E3"}; 
{\ar@(ul,ur) "F3"; "F3"}; 
{\ar@(ul,ur) "G3"; "G3"}; 
{\ar@(dr,dl) "A2"; "A2"}; 
{\ar@(dr,dl) "B2"; "B2"}; 
{\ar@(dr,dl) "C2"; "C2"}; 
{\ar@(dr,dl) "D2"; "D2"}; 
{\ar@(dr,dl) "E2"; "E2"}; 
{\ar@(dr,dl) "F2"; "F2"}; 
{\ar@(dl,dr) "G2"; "G2"}; 
\endxy
\]
\caption{State space of a bipartite graph}
\label{fig:bipartitestates}
\end{figure}

For a bipartite graph $K_{m_1,m_2}$ with $m_1 \neq m_2$, like for star graphs, the 
state space reduces to the product $\cals = \{0,1,\dots,m_1\} \times \{0,1,\dots,m_2\}$ 
where each ordered pair $(i,j) \in \cals$ indicates that there are $i$ vertices 
in $V_1$ and $j$ vertices in $V_2$ inhabited by mutant individuals. The 
evolutionary dynamics is described by the system of linear equations
\[
  P_{i,j}\rar \bigl(\Phi_{i+1,j} - \Phi_{i,j}\bigr) + P_{i,j}\lar \bigl(\Phi_{i-1,j} - \Phi_{i,j}\bigr) + P_{i,j}\uar \bigl(\Phi_{i,j+1} - \Phi_{i,j}\bigr) +  
  P_{i,j}\dar \bigl(\Phi_{i,j-1} - \Phi_{i,j}\bigr) = 0,
\]
under the boundary conditions $\Phi_{0,0} = 0$ and $\Phi_{m_1,m_2} = 1$, where 
the transition probabilities are given by 
\begin{align*}
  P_{i,j}\rar &= \Pp[ X_{n+1} = (i+1,j) | X_n = (i,j) ] = 
                    \frac{rj}{r(i+j)+N-(i+j)}\cdot \frac{m_1-i}{m_1}  \\
  P_{i,j}\lar &= \Pp[ X_{n+1} = (i-1,j) | X_n = (i,j) ] = 
                    \frac{m_2-j}{r(i+j)+N-(i+j)}\cdot \frac{i}{m_1}   \\
  P_{i,j}\uar &= \Pp[ X_{n+1}= (i,j+1)  | X_n = (i,j) ] = 
                    \frac{ri}{r(i+j)+N-(i+j)}\cdot \frac{m_2-j}{m_2}  \\
  P_{i,j}\dar &= \Pp[ X_{n+1}= (i,j-1)  | X_n = (i,j) ] = 
                    \frac{m_1-i}{r(i+j)+N-(i+j)}\cdot \frac{j}{m_2} 
\end{align*}
and $N = m_1 + m_2$. The subscript $(i,j)$ denote the initial state, 
while the arrows $\rar$, $\lar$, $\uar$, and $\dar$ are guidelines indicating the 
the direction of corresponding edge for the directed graph structure on the state space (so that the next state is $(i+1,j)$, $(i-1,j)$, $(i,j+1)$, or $(i,j-1)$ respectively), see Figure~\ref{fig:bipartitestates}.
By applying the loop-erasing method, we obtain the 
following new transition probabilities: 
\[
  \hat{P}_{i,m_2}\rar  = \frac{rm_2}{m_1+rm_2} 
  \qquad\text{and}\qquad
  \hat{P}_{i,m_2}\dar  = \frac{m_1}{m_1+rm_2}
\]
for the state $(i,m_2)$ and 
\begin{align}
  \hat{P}_{i,j}\rar &= \frac{r(m_1-i)jm_2}{(m_1 + rm_2)(m_1-i)j + i(m_2-j)(rm_1+m_2)}
      \label{rar} \\
  \hat{P}_{i,j}\lar &= \frac{im_2(m_2-j)}{(m_1 + rm_2)(m_1-i)j + i(m_2-j)(rm_1+m_2)} 
      \label{lar} \\
  \hat{P}_{i,j}\uar &= \frac{rim_1(m_2-j)}{(m_1 + rm_2)(m_1-i)j + i(m_2-j)(rm_1+m_2)} 
      \label{uar} \\
  \hat{P}_{i,j}\dar &= \frac{m_1(m_1-i)j}{(m_1 + rm_2)(m_1-i)j + i(m_2-j)(rm_1+m_2)} 
      \label{dar}
\end{align} 
for the state $(i,j)$ with $0 < j < m_2$. 
Like for star graphs, all the symmetries in complete bipartite graphs has been 
used to reduces the state space of the evolutionary process to the vertex set 
of the directed graph $\G$ described in Figure~\ref{fig:bipartitestates}. But 
neither this reduced process 
(random walk on $\G$), nor the process obtained by loop-erasing 
(random walk on the directed graph $\hat{\G}$ obtained by 
suppressing every loop connecting a non-absorbing state with itself) admit 
global symmetries. Note that states $(i,m_2)$ in the last row are never related 
to states $(i,j)$ with $1 \leq j < m_2$ by automorphisms of the weighted 
directed graphs $\G$ and $\hat{\G}$. Nevertheless, we prove that 
the map 
sending the state $(i,j)$ to the state $(i,j-1)$ with $1 <  j \leq m_2$ becomes 
more and more close to a symmetry of the Embedded Markov chain when 
$m_1 \to +\infty$. Using the 
previous calculation of the asymptotic fixation probability for a star graph, 
we obtain the following theorem, that is also illustrated numerically in 
Figure~\ref{fig:fixationbipartite}.

\begin{theorem}
  Let $\Phi_A(K_{m_1,m_2},r)$ be the average fixation probability of a single mutant individual 
  having a neutral or advantageous allele $A$ with fitness $r \geq 1$ in a Moran process on 
  a complete bipartite graph $K_{m_1,m_2}$. Then 
  \[
    \lim_{m_1 \to +\infty} \Phi_A(K_{m_1,m_2},r) 
                      = \lim_{m \to +\infty} \Phi_A(K_m,r^2) = 1-\frac1{r^2}
  \]
  where $\Phi_A(K_m,r^2)$ is the average fixation probability of a single mutant individual 
  having a neutral or advantageous allele $A$ with fitness $r^2 \geq 1$ in the classical 
  Moran process on $K_m$. 
\end{theorem}
\begin{proof}
  We start by observing that, according to~\eqref{rar}, we have:
  \[
    \hat{P}_{0,j}\rar 
    =  \frac{rm_2}{m_1 + rm_2}  
  \]
  for $1 \leq j \leq m_2$ and
  \begin{align*}
    \hat{P}_{i,j}\rar  - \hat{P}_{i,j-1}\rar & = 
    \frac{r(m_1-i)jm_2}{(m_1 + rm_2)(m_1-i)j + i(m_2-j)(rm_1+m_2)} \\
    & \quad - \frac{r(m_1-i)(j-1)m_2}{(m_1 + rm_2)(m_1-i)(j-1) + i(m_2-j+1)(rm_1+m_2)} \\
    & = \frac{rm_2}{m_1 + rm_2 + \dfrac{i}{m_1-i} \cdot \dfrac{m_2-j}{j} \, (rm_1+m_2)} \\
    & \quad - \frac{rm_2}{m_1 + rm_2 + \dfrac{i}{m_1-i} \cdot \dfrac{m_2-j+1}{j-1} \, (rm_1+m_2)} \\
    &< rm_2 \cdot \frac{\dfrac{i}{m_1-i} \, \Bigl( \dfrac{m_2-j+1}{j-1} - \dfrac{m_2-j}{j} \Bigr) \, (rm_1+m_2)}{(m_1 + rm_2)^2} \\
    &= rm_2 \cdot \frac{\dfrac{i}{m_1-i} \cdot \dfrac{m_2}{j(j-1)} \, (rm_1+m_2)}{(m_1 + rm_2)^2} \\
    &< rm_2 \cdot \frac{\dfrac{i}{m_1-i} \, m_2 \, (rm_1+m_2)}{(m_1 + rm_2)^2}
  \end{align*}
  for $2 \leq j \leq m_2$ and for $1 \leq i \leq m_1-1$.
  Assuming $m_1 \geq 2i$, we have $\frac{i}{m_1-i} \leq 1$
  and hence 
  \[
    \hat{P}_{i,j}\rar - \hat{P}_{i,j-1}\rar  
      <     rm_2 \cdot \frac{\dfrac{i}{m_1-i}\, m_2(rm_1+m_2)}{(m_1 + rm_2 )^2} 
      \leq  rm_2 \cdot \frac{m_2(rm_1+m_2)}{(m_1 + rm_2)^2}. \nonumber
  \]
  We deduce
  \begin{equation}
    \label{biparproof1i}
    \lim_{m_1 \to +\infty} \hat{P}_{i,j}\rar - \hat{P}_{i,j-1}\rar = 0
  \end{equation}
  for $2 \leq j \leq m_2$ and for $i \geq 1$. Similarly, using~\eqref{lar}, we 
  have $\hat{P}_{0,j}\lar = 0$ for $1 \leq j \leq m_2$, 
  $\hat{P}_{i,m_2}\lar = 0$ for $1 \leq i \leq m_1-1$, and
  \begin{align*}
    \hat{P}_{i,j}\lar & = \frac{im_2(m_2-j)}{(m_1 + rm_2)(m_1-i)j + i(m_2-j)(rm_1+m_2)} \\
    & = \frac{m_2}{(m_1 + rm_2)\, \dfrac{m_1-i}{i} \cdot \dfrac{j}{m_2-j} + rm_1+m_2} \\
    & < \frac{m_2}{rm_1+m_2}
  \end{align*}
  for $1 \leq j \leq m_1-1$. As before, it follows:
  \begin{equation}
    \label{biparproof2i}
    \lim_{m_1 \to +\infty} \hat{P}_{i,j}\lar =  \hat{P}_{i,m_2}\lar = 0
  \end{equation}
  for $1 \leq j\leq m_2$ and for each $i \geq 1$. Next, using~\eqref{uar}, we 
  have $\hat{P}_{0,j}\uar = 0$ for $1 \leq j \leq m_2$, 
  $\hat{P}_{i,m_2}\uar = 0$  for $1 \leq i \leq m_1-1$, and
  \begin{align*}
    \hat{P}_{i,j}\uar & = \frac{rim_1(m_2-j)}{(m_1 + rm_2)(m_1-i)j + i(m_2-j)(rm_1+m_2)} \\ 
    & = \frac{ri}{(m_1 + rm_2) \,\dfrac{m_1-i}{m_1} \cdot \dfrac{j}{m_2-j} + \dfrac{i}{m_1} (rm_1+m_2)} \\
    & < \frac{ri}{(m_1 + rm_2) \,\dfrac{m_1-i}{m_1} \cdot \dfrac{j}{m_2-j}} 
  \end{align*}
  for $1 \leq j \leq m_1-1$. Since $\frac12 \leq \frac{m_1 -i}{m_1}$ and 
  $\frac1{m_2} < \frac{j}{m_2-j}$ when $m_1 \geq 2i$ and $j \leq m_2-1$, we have 
  \[
    \hat{P}_{i,j}\uar < \frac{2m_2ri}{(m_1 + rm_2)}
  \]
  and therefore 
  \begin{equation}
    \label{biparproof3i}
    \lim_{m_1 \to +\infty} \hat{P}_{i,j}\uar = 0
  \end{equation}
  for $1 \leq j \leq m_2$ and for  $i \geq 1$. Finally, we have:
  \[
    \lim_{m_1 \to +\infty} \hat{P}_{i,j}\dar - \hat{P}_{i,j-1}\dar
      = \lim_{m_1 \to +\infty} \hat{P}_{i,j}\rar - \hat{P}_{i,j-1}\rar 
        + \hat{P}_{i,j}\lar - \hat{P}_{i,j-1}\lar
          + \hat{P}_{i,j}\uar - \hat{P}_{i,j-1}\uar = 0 
  \]
  from \eqref{biparproof1i}, \eqref{biparproof2i} and \eqref{biparproof3i}. 
  Arguing inductively on the integer $i \geq 1$, this implies that the Moran 
  process on the bipartite graph $K_{m_1,m_2}$ reduces asymptotically to the 
  Moran process on the star $K_{m_1,1}$ when $m_1 \to +\infty$, and hence 
  \[
    \lim_{m_1 \to +\infty} \Phi_A(K_{m_1,m_2},r) - \Phi_A(r,K_{m_1,1},r) = 0,
  \]
  that proves the theorem.
\end{proof}

\section{Numerical experiments in complex networks}
\label{sec:experiments}

Proposition~\ref{thexabtime} says that the expected number of steps until 
absorption in the loop-erased Markov chain is smaller or equal than that in 
the standard one. At first glance, this seems to imply that Monte Carlo method 
on the EMC (\emph{EMC method} from now on) will stop before Monte Carlo on 
the standard chain, (\emph{Standard Monte Carlo} or \emph{SMC method} from now 
on), but there is a subtle difference between what the method does theoretically and what the computer actually does.

First of all, we need to construct a weighted directed graph $\G$ having $2^n$ states. It is almost always unfeasible when $n$ is relatively large, but it is easier for highly symmetric graphs.
So simulations reproduce how individuals randomly spawn and die. In the SMC method, at each step, the 
chance of selecting a mutant individual for reproduction is proportional to the 
fitness $r\geq1$. This uniformity allows us to update the new transition 
probabilities in constant time. 
However, in the EMC method the probability of choosing each individual for reproduction depends not only on its fitness but also on the fitness of its neighbours. More precisely, the probability that a particular mutant individual $v$ leaves offspring at a particular time is proportional to the number of resident neighbours of $v$ at that time. Similarly, if $v$ is a resident individual, the probability of choosing it for reproduction is proportional to the number of mutant neighbours. 
Thus, if $w$ is the neighbour of $v$ chosen to die, the EMC method needs to update the transition probabilities of each neighbour of $w$.
On some graphs, this may lead to longer computation times. 
\medskip 

We compared the amount of time it takes to end the simulations for the two 
methods in a series of well-known complex network models. All simulations were 
done on a computer running MacOS~X~10.9.3 with a quad-core~i5 at $2.5$GHz and 
$8$Gb of RAM. Graph construction and manipulation was done in
Sage/NetworkX~\cite{HSS,sage}, but the simulation routines were written in C.
\begin{figure}
  \centering
  \includegraphics[width=0.95\textwidth]{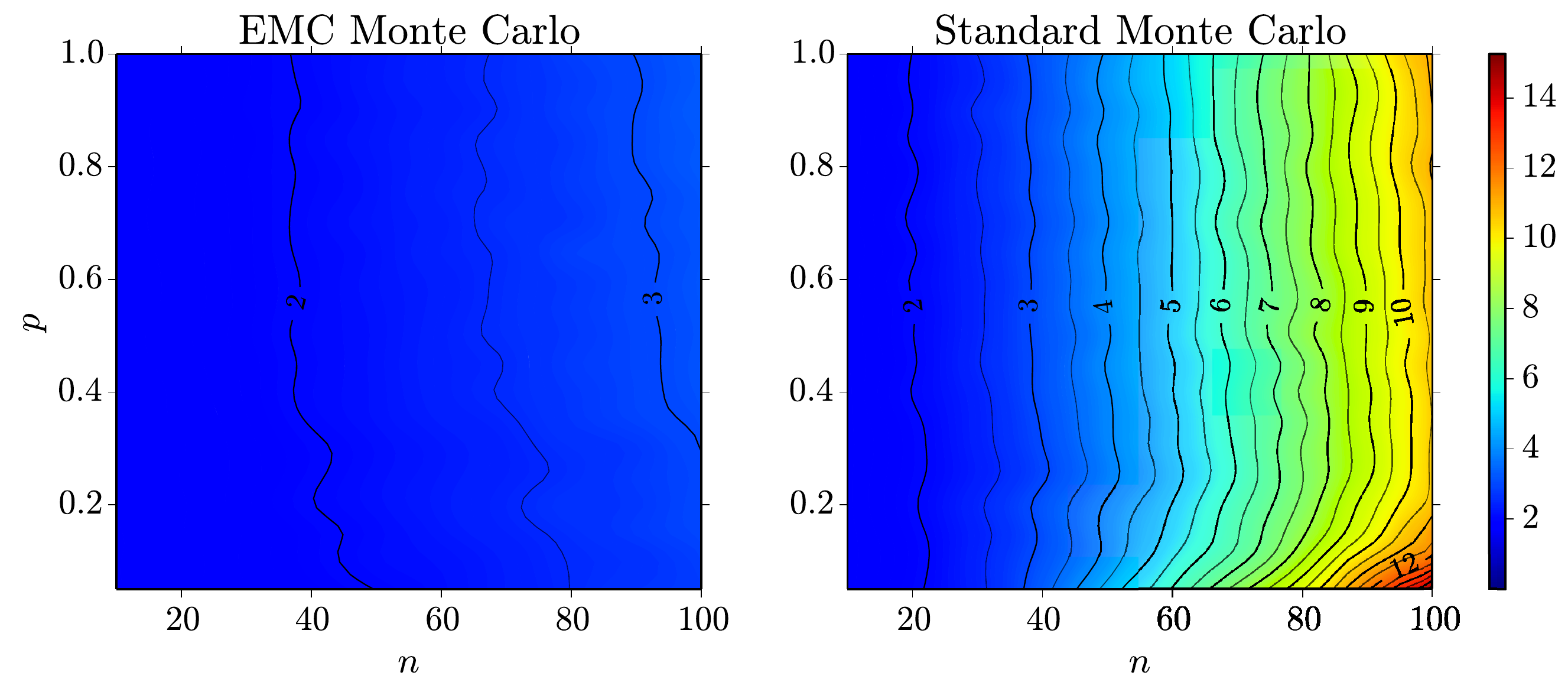}
  \caption{Average computation times (in seconds) for Moran processes on 
    small-world 
    networks (Watts-Strogatz $\beta$-model) using EMC and SMC methods for all 
    $r$ going from $0$ to $10$ with step size of $0.25$.}
  \label{fig:small_worlds}
\end{figure}
\subsection*{Small-world networks} 

Small-world networks were introduced in~\cite{WS} as a family of random 
graphs with some properties of real networks. The construction is as 
follows: consider a circular graph of order $n$ and connect the $k$ nearest 
neighbours. Now, each edge $uv$ in the previous graph can be replaced with 
another edge $uw$ with probability $p$. The resulting graph may be disconnected.
\medskip

We did the following experiment to test the speed of the two methods: 
Fixed $k=4$, for any $n\in\{10,20,\ldots,100\}$ and $p\in\{0,0.05,\ldots,1\}$, 
\begin{list}{$\bullet$}{\leftmargin=1em}

\item we construct $10$ random graphs with parameters $n$ and $p$;

\item for each of these graphs, we compute the average fixation probability 
  using both methods $3$ times with $1000$ trials for every fitness $r$ varying 
  from $0$ to $10$ with step size of $0.25$.
  
\end{list}
Averaging the $30$ running times of each method we get an \emph{average computation time} for both algorithms on the family of small-world networks with the prescribed parameters. The results are shown in Figure~\ref{fig:small_worlds}. As can be seen, the EMC method performs better than the SMC method on this family of networks. 

\begin{figure}
  \centering
  \includegraphics[width=0.95\textwidth]{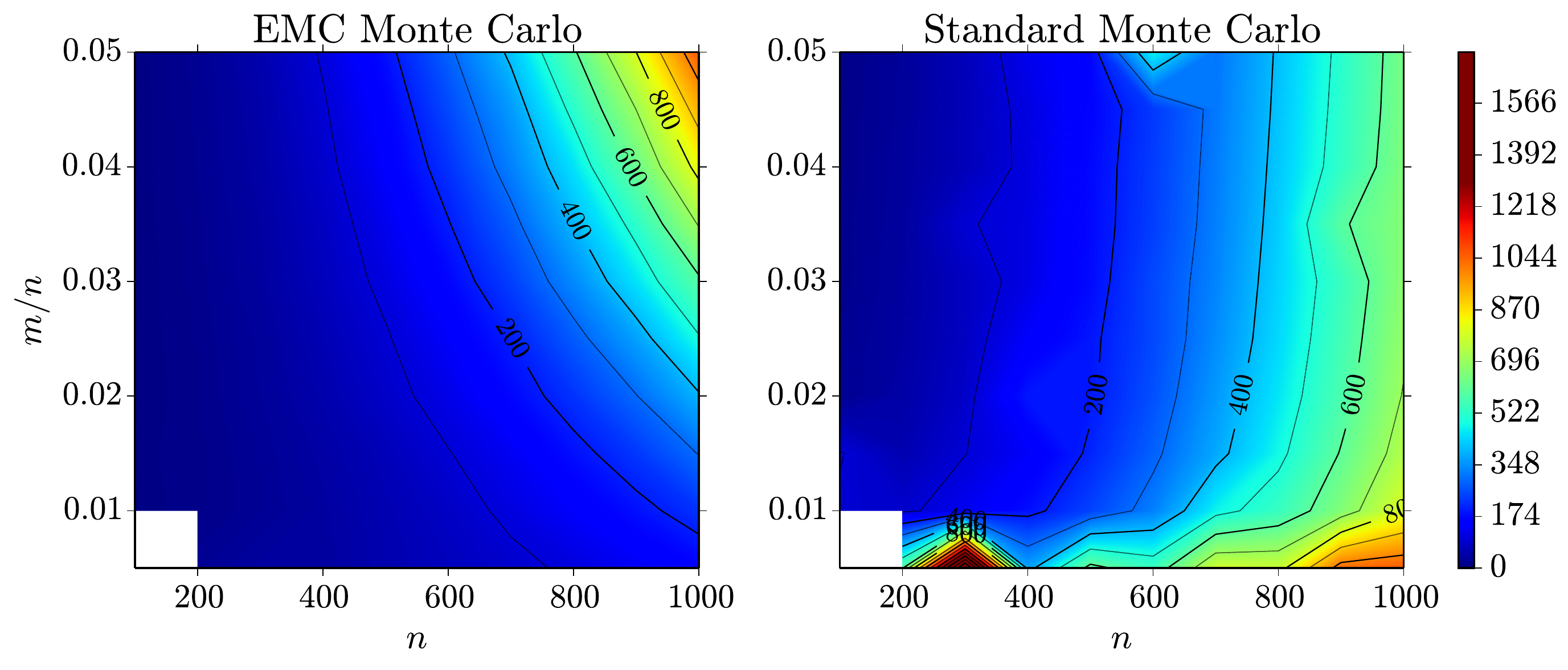}
  \caption{Average computation times (in seconds) for Moran processes on 
    scale-free networks (Barab\'asi-Albert preferential attachment model) using 
    EMC and SMC methods for 
    $r$ going from $0$ to $10$ with step size of $0.25$.}
  \label{fig:BA}
\end{figure}

\subsection*{Scale-free networks}

The previous family of graphs lacked a fairly common property of real networks, 
namely a power-law degree distribution. In~\cite{BA} the preferential 
attachment model was developed to solve the shortcomings of previous models. 
Start with $m$ vertices connected in no way. Then one single 
vertex is added and connected to the initial vertices to obtain a star. At 
successive steps, another single vertex is added and connected to $m$ of the 
previous vertices with probability `proportional' to the degree. After $n-m$ 
steps, the graph has $n$ vertices and $(n-m)m$ edges.
In the real world, one expect to have small $m$ compared to $n$ as the global 
population is large and people known just a very small portion of the 
population.
\medskip

We ran a similar experiment as for small-world networks, using both methods $3$ 
times for $10$ random graphs with $1000$ trials for every fitness $r$  varying 
from $0$ to $10$ in $40$ evenly disposed steps, but new relevant parameters are 
now the order of the graph $n\in\{100,200,\ldots,1000\}$ and the ratio 
$m/n\in\{0.005,0.01,\ldots,0.05\}$. Thus, a point $(n,m/n)$ in the 
plot corresponds to the average computation time on the random family with 
parameters $n$ and $m = \lfloor n\cdot m/n\rfloor$. 
The size of the population $n$ has been multiplied by $10$ with respect to the 
size of the small-world networks in the previous sample because we need to 
consider a population with $n\leq100$ individuals if $m/n=0.01$ and $n\geq200$ 
individuals if $m/n = 0.005$.
The result of the simulation can be seen in Figure~\ref{fig:BA}. 
The SMC method performance is specially bad on graphs obtained by 
Barab\'asi-Albert preferential attachment with $m=1$, whereas EMC method 
performs badly on the models with `large' $m$ where high degree vertices 
appear. Star graphs can be interpreted as graphs obtained by Barab\'asi-Albert 
preferential attachment in a single step. According to this interpretation, SMC 
method should improve the average computation times obtained by the EMC method. 
In Table~\ref{table:EMC-star-Knm}, we compare these times on some star and 
complete bipartite graphs with the same number of trials and values of the 
fitness $r$. 
%
\begin{table}
  \centering
  \begin{tabular}{cccc}
    \toprule
    Graph      &     EMC time &     SMC time & SMC/EMC \\ 
    \midrule
    $K_{1,10}  $ & \invi{147}0.98 & \invi{136}0.97 &  0.99 \\ 
    $K_{1,50}  $ & \invi{14}23.79 & \invi{13}30.04 &  1.26 \\ 
    $K_{1,100} $ & \invi{1}181.48 & \invi{1}190.43 &  1.05 \\ 
    $K_{1,200} $ &        1471.91 &        1369.67 &  0.93 \\ 
    \midrule
    $K_{2,10}  $ & \invi{147}0.81 & \invi{136}0.88 &  1.09 \\ 
    $K_{2,50}  $ & \invi{14}12.52 & \invi{13}16.17 &  1.29 \\ 
    $K_{2,100} $ & \invi{14}95.10 & \invi{13}98.49 &  1.04 \\ 
    $K_{2,200} $ & \invi{1}753.86 & \invi{1}694.88 &  0.92 \\ 
    \midrule
    $K_{10,10} $ & \invi{147}0.91 & \invi{136}0.90 &  0.99 \\ 
    $K_{10,50} $ & \invi{147}4.72 & \invi{136}4.53 &  0.96 \\ 
    $K_{10,100}$ & \invi{14}29.68 & \invi{13}23.68 &  0.80 \\ 
    $K_{10,200}$ & \invi{1}217.40 & \invi{1}154.32 &  0.71 \\
    \bottomrule 
  \end{tabular}

  \caption{Average computation times (in seconds) for Moran processes on the 
    star graphs of orders $11$, $51$, $101$ and $201$, and on the 
    complete bipartite graphs $K_{2,10}$, $K_{2,50}$, $K_{2,100}$, $K_{2,200}$, 
    $K_{10,10}$, $K_{10,50}$, $K_{10,100}$ and $K_{10,200}$. All simulations 
    with $1000$ trials and $r$ going from $0$ to $10$ with step size of $0.25$. 
    The last column shows how many times faster EMC is than the SMC method.}
  \label{table:EMC-star-Knm}
\end{table}

\subsection*{Hierarchical networks}
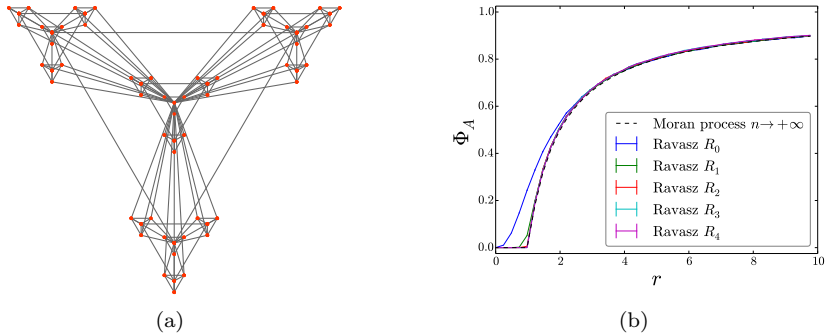
\begin{figure}[t!]
  \centering
  \subfigure[]{
  \begin{tikzpicture}[line cap=round,line join=round,>=triangle 45,x=1.0cm,y=1.0cm,scale=.15]
    \clip(-17,-17) rectangle (17,10);
    \draw [color=zzzzzz] (0,-1)-- (0,0);
    \draw [color=zzzzzz] (0.87,0.5)-- (0,0);
    \draw [color=zzzzzz] (-0.87,0.5)-- (0,0);
    \draw [color=zzzzzz] (-0.87,0.5)-- (0.87,0.5);
    \draw [color=zzzzzz] (0,-1)-- (0.87,0.5);
    \draw [color=zzzzzz] (0,-1)-- (-0.87,0.5);
    \draw [color=zzzzzz] (-0.87,-2.86)-- (0,-3.36);
    \draw [color=zzzzzz] (0,-4.36)-- (0,-3.36);
    \draw [color=zzzzzz] (0.87,-2.86)-- (0,-3.36);
    \draw [color=zzzzzz] (0.87,-2.86)-- (0,-4.36);
    \draw [color=zzzzzz] (-0.87,-2.86)-- (0,-4.36);
    \draw [color=zzzzzz] (-0.87,-2.86)-- (0.87,-2.86);
    \draw [color=zzzzzz] (2.91,0.68)-- (2.91,1.68);
    \draw [color=zzzzzz] (3.78,2.18)-- (2.91,1.68);
    \draw [color=zzzzzz] (2.04,2.18)-- (2.91,1.68);
    \draw [color=zzzzzz] (2.04,2.18)-- (3.78,2.18);
    \draw [color=zzzzzz] (2.91,0.68)-- (3.78,2.18);
    \draw [color=zzzzzz] (2.91,0.68)-- (2.04,2.18);
    \draw [color=zzzzzz] (-2.04,2.18)-- (-2.91,1.68);
    \draw [color=zzzzzz] (-3.78,2.18)-- (-2.91,1.68);
    \draw [color=zzzzzz] (-2.91,0.68)-- (-2.91,1.68);
    \draw [color=zzzzzz] (-2.91,0.68)-- (-3.78,2.18);
    \draw [color=zzzzzz] (-2.04,2.18)-- (-3.78,2.18);
    \draw [color=zzzzzz] (-2.04,2.18)-- (-2.91,0.68);
    \draw [color=zzzzzz] (0,0)-- (2.04,2.18);
    \draw [color=zzzzzz] (0,0)-- (2.91,0.68);
    \draw [color=zzzzzz] (0,0)-- (0.87,-2.86);
    \draw [color=zzzzzz] (0,0)-- (-0.87,-2.86);
    \draw [color=zzzzzz] (0,0)-- (-2.91,0.68);
    \draw [color=zzzzzz] (0,0)-- (-2.04,2.18);
    \draw [color=zzzzzz] (-2.91,1.68)-- (2.91,1.68);
    \draw [color=zzzzzz] (-2.91,1.68)-- (0,-3.36);
    \draw [color=zzzzzz] (0,-3.36)-- (2.91,1.68);
    \draw [color=zzzzzz] (0,0)-- (3.78,2.18);
    \draw [color=zzzzzz] (0,0)-- (0,-4.36);
    \draw [color=zzzzzz] (0,0)-- (-3.78,2.18);
    \draw [color=zzzzzz] (-0.87,-11.9)-- (-0.01,-12.4);
    \draw [color=zzzzzz] (-0.01,-13.4)-- (-0.01,-12.4);
    \draw [color=zzzzzz] (0.86,-11.9)-- (-0.01,-12.4);
    \draw [color=zzzzzz] (0.86,-11.9)-- (-0.01,-13.4);
    \draw [color=zzzzzz] (-0.87,-11.9)-- (-0.01,-13.4);
    \draw [color=zzzzzz] (-0.87,-11.9)-- (0.86,-11.9);
    \draw [color=zzzzzz] (-2.05,-10.22)-- (-2.91,-10.72);
    \draw [color=zzzzzz] (-3.78,-10.22)-- (-2.91,-10.72);
    \draw [color=zzzzzz] (-2.91,-11.72)-- (-2.91,-10.72);
    \draw [color=zzzzzz] (-2.91,-11.72)-- (-3.78,-10.22);
    \draw [color=zzzzzz] (-2.05,-10.22)-- (-3.78,-10.22);
    \draw [color=zzzzzz] (-2.05,-10.22)-- (-2.91,-11.72);
    \draw [color=zzzzzz] (-0.87,-15.26)-- (-0.01,-15.76);
    \draw [color=zzzzzz] (-0.01,-16.76)-- (-0.01,-15.76);
    \draw [color=zzzzzz] (0.86,-15.26)-- (-0.01,-15.76);
    \draw [color=zzzzzz] (0.86,-15.26)-- (-0.01,-16.76);
    \draw [color=zzzzzz] (-0.87,-15.26)-- (-0.01,-16.76);
    \draw [color=zzzzzz] (-0.87,-15.26)-- (0.86,-15.26);
    \draw [color=zzzzzz] (2.9,-11.72)-- (2.9,-10.72);
    \draw [color=zzzzzz] (3.77,-10.22)-- (2.9,-10.72);
    \draw [color=zzzzzz] (2.04,-10.22)-- (2.9,-10.72);
    \draw [color=zzzzzz] (2.04,-10.22)-- (3.77,-10.22);
    \draw [color=zzzzzz] (2.9,-11.72)-- (3.77,-10.22);
    \draw [color=zzzzzz] (2.9,-11.72)-- (2.04,-10.22);
    \draw [color=zzzzzz] (-0.01,-12.4)-- (0.86,-15.26);
    \draw [color=zzzzzz] (-0.01,-12.4)-- (-0.87,-15.26);
    \draw [color=zzzzzz] (-0.01,-12.4)-- (-2.91,-11.72);
    \draw [color=zzzzzz] (-0.01,-12.4)-- (-2.05,-10.22);
    \draw [color=zzzzzz] (-0.01,-12.4)-- (2.04,-10.22);
    \draw [color=zzzzzz] (-0.01,-12.4)-- (2.9,-11.72);
    \draw [color=zzzzzz] (2.9,-10.72)-- (-0.01,-15.76);
    \draw [color=zzzzzz] (2.9,-10.72)-- (-2.91,-10.72);
    \draw [color=zzzzzz] (-2.91,-10.72)-- (-0.01,-15.76);
    \draw [color=zzzzzz] (-0.01,-12.4)-- (-0.01,-16.76);
    \draw [color=zzzzzz] (-0.01,-12.4)-- (-3.78,-10.22);
    \draw [color=zzzzzz] (-0.01,-12.4)-- (3.77,-10.22);
    \draw [color=zzzzzz] (10.74,5.19)-- (10.74,6.19);
    \draw [color=zzzzzz] (11.6,6.69)-- (10.74,6.19);
    \draw [color=zzzzzz] (9.87,6.69)-- (10.74,6.19);
    \draw [color=zzzzzz] (9.87,6.69)-- (11.6,6.69);
    \draw [color=zzzzzz] (10.74,5.19)-- (11.6,6.69);
    \draw [color=zzzzzz] (10.74,5.19)-- (9.87,6.69);
    \draw [color=zzzzzz] (9.87,3.33)-- (10.74,2.83);
    \draw [color=zzzzzz] (10.74,1.83)-- (10.74,2.83);
    \draw [color=zzzzzz] (11.6,3.33)-- (10.74,2.83);
    \draw [color=zzzzzz] (11.6,3.33)-- (10.74,1.83);
    \draw [color=zzzzzz] (9.87,3.33)-- (10.74,1.83);
    \draw [color=zzzzzz] (9.87,3.33)-- (11.6,3.33);
    \draw [color=zzzzzz] (13.65,6.87)-- (13.65,7.87);
    \draw [color=zzzzzz] (14.51,8.37)-- (13.65,7.87);
    \draw [color=zzzzzz] (12.78,8.37)-- (13.65,7.87);
    \draw [color=zzzzzz] (12.78,8.37)-- (14.51,8.37);
    \draw [color=zzzzzz] (13.65,6.87)-- (14.51,8.37);
    \draw [color=zzzzzz] (13.65,6.87)-- (12.78,8.37);
    \draw [color=zzzzzz] (8.69,8.37)-- (7.83,7.87);
    \draw [color=zzzzzz] (6.96,8.37)-- (7.83,7.87);
    \draw [color=zzzzzz] (7.83,6.87)-- (7.83,7.87);
    \draw [color=zzzzzz] (7.83,6.87)-- (6.96,8.37);
    \draw [color=zzzzzz] (8.69,8.37)-- (6.96,8.37);
    \draw [color=zzzzzz] (8.69,8.37)-- (7.83,6.87);
    \draw [color=zzzzzz] (10.74,6.19)-- (12.78,8.37);
    \draw [color=zzzzzz] (10.74,6.19)-- (13.65,6.87);
    \draw [color=zzzzzz] (10.74,6.19)-- (11.6,3.33);
    \draw [color=zzzzzz] (10.74,6.19)-- (9.87,3.33);
    \draw [color=zzzzzz] (10.74,6.19)-- (7.83,6.87);
    \draw [color=zzzzzz] (10.74,6.19)-- (8.69,8.37);
    \draw [color=zzzzzz] (7.83,7.87)-- (13.65,7.87);
    \draw [color=zzzzzz] (7.83,7.87)-- (10.74,2.83);
    \draw [color=zzzzzz] (10.74,2.83)-- (13.65,7.87);
    \draw [color=zzzzzz] (10.74,6.19)-- (14.51,8.37);
    \draw [color=zzzzzz] (10.74,6.19)-- (10.74,1.83);
    \draw [color=zzzzzz] (10.74,6.19)-- (6.96,8.37);
    \draw [color=zzzzzz] (-9.87,6.7)-- (-10.73,6.2);
    \draw [color=zzzzzz] (-11.6,6.7)-- (-10.73,6.2);
    \draw [color=zzzzzz] (-10.73,5.2)-- (-10.73,6.2);
    \draw [color=zzzzzz] (-10.73,5.2)-- (-11.6,6.7);
    \draw [color=zzzzzz] (-9.87,6.7)-- (-11.6,6.7);
    \draw [color=zzzzzz] (-9.87,6.7)-- (-10.73,5.2);
    \draw [color=zzzzzz] (-7.82,6.88)-- (-7.82,7.88);
    \draw [color=zzzzzz] (-6.96,8.38)-- (-7.82,7.88);
    \draw [color=zzzzzz] (-8.69,8.38)-- (-7.82,7.88);
    \draw [color=zzzzzz] (-8.69,8.38)-- (-6.96,8.38);
    \draw [color=zzzzzz] (-7.82,6.88)-- (-6.96,8.38);
    \draw [color=zzzzzz] (-7.82,6.88)-- (-8.69,8.38);
    \draw [color=zzzzzz] (-12.78,8.38)-- (-13.64,7.88);
    \draw [color=zzzzzz] (-14.51,8.38)-- (-13.64,7.88);
    \draw [color=zzzzzz] (-13.64,6.88)-- (-13.64,7.88);
    \draw [color=zzzzzz] (-13.64,6.88)-- (-14.51,8.38);
    \draw [color=zzzzzz] (-12.78,8.38)-- (-14.51,8.38);
    \draw [color=zzzzzz] (-12.78,8.38)-- (-13.64,6.88);
    \draw [color=zzzzzz] (-11.6,3.34)-- (-10.73,2.84);
    \draw [color=zzzzzz] (-10.73,1.84)-- (-10.73,2.84);
    \draw [color=zzzzzz] (-9.87,3.34)-- (-10.73,2.84);
    \draw [color=zzzzzz] (-9.87,3.34)-- (-10.73,1.84);
    \draw [color=zzzzzz] (-11.6,3.34)-- (-10.73,1.84);
    \draw [color=zzzzzz] (-11.6,3.34)-- (-9.87,3.34);
    \draw [color=zzzzzz] (-10.73,6.2)-- (-13.64,6.88);
    \draw [color=zzzzzz] (-10.73,6.2)-- (-12.78,8.38);
    \draw [color=zzzzzz] (-10.73,6.2)-- (-8.69,8.38);
    \draw [color=zzzzzz] (-10.73,6.2)-- (-7.82,6.88);
    \draw [color=zzzzzz] (-10.73,6.2)-- (-9.87,3.34);
    \draw [color=zzzzzz] (-10.73,6.2)-- (-11.6,3.34);
    \draw [color=zzzzzz] (-10.73,2.84)-- (-13.64,7.88);
    \draw [color=zzzzzz] (-10.73,2.84)-- (-7.82,7.88);
    \draw [color=zzzzzz] (-7.82,7.88)-- (-13.64,7.88);
    \draw [color=zzzzzz] (-10.73,6.2)-- (-14.51,8.38);
    \draw [color=zzzzzz] (-10.73,6.2)-- (-6.96,8.38);
    \draw [color=zzzzzz] (-10.73,6.2)-- (-10.73,1.84);
    \draw [color=zzzzzz] (0,0)-- (6.96,8.37);
    \draw [color=zzzzzz] (0,0)-- (8.69,8.37);
    \draw [color=zzzzzz] (0,0)-- (7.83,6.87);
    \draw [color=zzzzzz] (0,0)-- (12.78,8.37);
    \draw [color=zzzzzz] (0.87,0.5)-- (14.51,8.37);
    \draw [color=zzzzzz] (0,0)-- (13.65,6.87);
    \draw [color=zzzzzz] (0,0)-- (9.87,3.33);
    \draw [color=zzzzzz] (0,0)-- (11.6,3.33);
    \draw [color=zzzzzz] (0,0)-- (10.74,1.83);
    \draw [color=zzzzzz] (0,0)-- (3.77,-10.22);
    \draw [color=zzzzzz] (2.04,-10.22)-- (0,0);
    \draw [color=zzzzzz] (2.9,-11.72)-- (0,0);
    \draw [color=zzzzzz] (0,0)-- (-2.05,-10.22);
    \draw [color=zzzzzz] (0,0)-- (-3.78,-10.22);
    \draw [color=zzzzzz] (0,0)-- (-2.91,-11.72);
    \draw [color=zzzzzz] (0,0)-- (-0.87,-15.26);
    \draw [color=zzzzzz] (0,0)-- (-0.01,-16.76);
    \draw [color=zzzzzz] (0,0)-- (0.86,-15.26);
    \draw [color=zzzzzz] (0,0)-- (-6.96,8.38);
    \draw [color=zzzzzz] (0,0)-- (-7.82,6.88);
    \draw [color=zzzzzz] (0,0)-- (-8.69,8.38);
    \draw [color=zzzzzz] (0,0)-- (-10.73,1.84);
    \draw [color=zzzzzz] (0,0)-- (-9.87,3.34);
    \draw [color=zzzzzz] (0,0)-- (-11.6,3.34);
    \draw [color=zzzzzz] (0,0)-- (-14.51,8.38);
    \draw [color=zzzzzz] (0,0)-- (-12.78,8.38);
    \draw [color=zzzzzz] (0,0)-- (-13.64,6.88);
    \draw [color=zzzzzz] (-0.01,-12.4)-- (10.74,6.19);
    \draw [color=zzzzzz] (10.74,6.19)-- (-10.73,6.2);
    \draw [color=zzzzzz] (-10.73,6.2)-- (-0.01,-12.4);
    \begin{scriptsize}
      \fill [color=ffttqq] (0,-1) circle (4.5pt);
      \fill [color=ffttqq] (0,0) circle (4.5pt);
      \fill [color=ffttqq] (0.87,0.5) circle (4.5pt);
      \fill [color=ffttqq] (0,0) circle (4.5pt);
      \fill [color=ffttqq] (-0.87,0.5) circle (4.5pt);
      \fill [color=ffttqq] (0,0) circle (4.5pt);
      \fill [color=ffttqq] (0,-3.36) circle (4.5pt);
      \fill [color=ffttqq] (0,-3.36) circle (4.5pt);
      \fill [color=ffttqq] (0,-3.36) circle (4.5pt);
      \fill [color=ffttqq] (-0.87,-2.86) circle (4.5pt);
      \fill [color=ffttqq] (0,-3.36) circle (4.5pt);
      \fill [color=ffttqq] (0,-4.36) circle (4.5pt);
      \fill [color=ffttqq] (0,-3.36) circle (4.5pt);
      \fill [color=ffttqq] (0.87,-2.86) circle (4.5pt);
      \fill [color=ffttqq] (0,-3.36) circle (4.5pt);
      \fill [color=ffttqq] (0.87,-2.86) circle (4.5pt);
      \fill [color=ffttqq] (0,-4.36) circle (4.5pt);
      \fill [color=ffttqq] (-0.87,-2.86) circle (4.5pt);
      \fill [color=ffttqq] (0,-4.36) circle (4.5pt);
      \fill [color=ffttqq] (-0.87,-2.86) circle (4.5pt);
      \fill [color=ffttqq] (0.87,-2.86) circle (4.5pt);
      \fill [color=ffttqq] (2.91,0.68) circle (4.5pt);
      \fill [color=ffttqq] (2.91,1.68) circle (4.5pt);
      \fill [color=ffttqq] (3.78,2.18) circle (4.5pt);
      \fill [color=ffttqq] (2.91,1.68) circle (4.5pt);
      \fill [color=ffttqq] (2.04,2.18) circle (4.5pt);
      \fill [color=ffttqq] (2.91,1.68) circle (4.5pt);
      \fill [color=ffttqq] (2.04,2.18) circle (4.5pt);
      \fill [color=ffttqq] (3.78,2.18) circle (4.5pt);
      \fill [color=ffttqq] (2.91,0.68) circle (4.5pt);
      \fill [color=ffttqq] (3.78,2.18) circle (4.5pt);
      \fill [color=ffttqq] (2.91,0.68) circle (4.5pt);
      \fill [color=ffttqq] (2.04,2.18) circle (4.5pt);
      \fill [color=ffttqq] (2.91,1.68) circle (4.5pt);
      \fill [color=ffttqq] (2.91,1.68) circle (4.5pt);
      \fill [color=ffttqq] (2.91,1.68) circle (4.5pt);
      \fill [color=ffttqq] (-2.04,2.18) circle (4.5pt);
      \fill [color=ffttqq] (-2.91,1.68) circle (4.5pt);
      \fill [color=ffttqq] (-3.78,2.18) circle (4.5pt);
      \fill [color=ffttqq] (-2.91,1.68) circle (4.5pt);
      \fill [color=ffttqq] (-2.91,0.68) circle (4.5pt);
      \fill [color=ffttqq] (-2.91,1.68) circle (4.5pt);
      \fill [color=ffttqq] (-2.91,0.68) circle (4.5pt);
      \fill [color=ffttqq] (-3.78,2.18) circle (4.5pt);
      \fill [color=ffttqq] (-2.04,2.18) circle (4.5pt);
      \fill [color=ffttqq] (-3.78,2.18) circle (4.5pt);
      \fill [color=ffttqq] (-2.04,2.18) circle (4.5pt);
      \fill [color=ffttqq] (-2.91,0.68) circle (4.5pt);
      \fill [color=ffttqq] (-2.91,1.68) circle (4.5pt);
      \fill [color=ffttqq] (-2.91,1.68) circle (4.5pt);
      \fill [color=ffttqq] (-2.91,1.68) circle (4.5pt);
      \fill [color=ffttqq] (-0.01,-12.4) circle (4.5pt);
      \fill [color=ffttqq] (-0.01,-12.4) circle (4.5pt);
      \fill [color=ffttqq] (-0.01,-12.4) circle (4.5pt);
      \fill [color=ffttqq] (-0.01,-12.4) circle (4.5pt);
      \fill [color=ffttqq] (-0.01,-12.4) circle (4.5pt);
      \fill [color=ffttqq] (-0.01,-12.4) circle (4.5pt);
      \fill [color=ffttqq] (-0.01,-12.4) circle (4.5pt);
      \fill [color=ffttqq] (-0.01,-12.4) circle (4.5pt);
      \fill [color=ffttqq] (-0.01,-12.4) circle (4.5pt);
      \fill [color=ffttqq] (-0.01,-12.4) circle (4.5pt);
      \fill [color=ffttqq] (-0.01,-12.4) circle (4.5pt);
      \fill [color=ffttqq] (-0.01,-12.4) circle (4.5pt);
      \fill [color=ffttqq] (-0.87,-11.9) circle (4.5pt);
      \fill [color=ffttqq] (-0.01,-12.4) circle (4.5pt);
      \fill [color=ffttqq] (-0.01,-13.4) circle (4.5pt);
      \fill [color=ffttqq] (-0.01,-12.4) circle (4.5pt);
      \fill [color=ffttqq] (0.86,-11.9) circle (4.5pt);
      \fill [color=ffttqq] (-0.01,-12.4) circle (4.5pt);
      \fill [color=ffttqq] (0.86,-11.9) circle (4.5pt);
      \fill [color=ffttqq] (-0.01,-13.4) circle (4.5pt);
      \fill [color=ffttqq] (-0.87,-11.9) circle (4.5pt);
      \fill [color=ffttqq] (-0.01,-13.4) circle (4.5pt);
      \fill [color=ffttqq] (-0.87,-11.9) circle (4.5pt);
      \fill [color=ffttqq] (0.86,-11.9) circle (4.5pt);
      \fill [color=ffttqq] (-2.91,-10.72) circle (4.5pt);
      \fill [color=ffttqq] (-3.78,-10.22) circle (4.5pt);
      \fill [color=ffttqq] (-2.91,-10.72) circle (4.5pt);
      \fill [color=ffttqq] (-2.91,-11.72) circle (4.5pt);
      \fill [color=ffttqq] (-2.91,-10.72) circle (4.5pt);
      \fill [color=ffttqq] (-2.91,-11.72) circle (4.5pt);
      \fill [color=ffttqq] (-3.78,-10.22) circle (4.5pt);
      \fill [color=ffttqq] (-3.78,-10.22) circle (4.5pt);
      \fill [color=ffttqq] (-2.05,-10.22) circle (4.5pt);
      \fill [color=ffttqq] (-2.91,-11.72) circle (4.5pt);
      \fill [color=ffttqq] (-0.87,-15.26) circle (4.5pt);
      \fill [color=ffttqq] (-0.01,-15.76) circle (4.5pt);
      \fill [color=ffttqq] (-0.01,-16.76) circle (4.5pt);
      \fill [color=ffttqq] (-0.01,-15.76) circle (4.5pt);
      \fill [color=ffttqq] (0.86,-15.26) circle (4.5pt);
      \fill [color=ffttqq] (-0.01,-15.76) circle (4.5pt);
      \fill [color=ffttqq] (0.86,-15.26) circle (4.5pt);
      \fill [color=ffttqq] (-0.01,-16.76) circle (4.5pt);
      \fill [color=ffttqq] (-0.87,-15.26) circle (4.5pt);
      \fill [color=ffttqq] (-0.01,-16.76) circle (4.5pt);
      \fill [color=ffttqq] (-0.87,-15.26) circle (4.5pt);
      \fill [color=ffttqq] (0.86,-15.26) circle (4.5pt);
      \fill [color=ffttqq] (2.9,-11.72) circle (4.5pt);
      \fill [color=ffttqq] (2.9,-10.72) circle (4.5pt);
      \fill [color=ffttqq] (3.77,-10.22) circle (4.5pt);
      \fill [color=ffttqq] (2.9,-10.72) circle (4.5pt);
      \fill [color=ffttqq] (2.04,-10.22) circle (4.5pt);
      \fill [color=ffttqq] (2.9,-10.72) circle (4.5pt);
      \fill [color=ffttqq] (2.04,-10.22) circle (4.5pt);
      \fill [color=ffttqq] (3.77,-10.22) circle (4.5pt);
      \fill [color=ffttqq] (2.9,-11.72) circle (4.5pt);
      \fill [color=ffttqq] (3.77,-10.22) circle (4.5pt);
      \fill [color=ffttqq] (2.9,-11.72) circle (4.5pt);
      \fill [color=ffttqq] (2.04,-10.22) circle (4.5pt);
      \fill [color=ffttqq] (-0.01,-12.4) circle (4.5pt);
      \fill [color=ffttqq] (0.86,-15.26) circle (4.5pt);
      \fill [color=ffttqq] (-0.01,-12.4) circle (4.5pt);
      \fill [color=ffttqq] (-0.87,-15.26) circle (4.5pt);
      \fill [color=ffttqq] (-0.01,-12.4) circle (4.5pt);
      \fill [color=ffttqq] (-2.91,-11.72) circle (4.5pt);
      \fill [color=ffttqq] (-0.01,-12.4) circle (4.5pt);
      \fill [color=ffttqq] (-2.05,-10.22) circle (4.5pt);
      \fill [color=ffttqq] (-0.01,-12.4) circle (4.5pt);
      \fill [color=ffttqq] (2.04,-10.22) circle (4.5pt);
      \fill [color=ffttqq] (-0.01,-12.4) circle (4.5pt);
      \fill [color=ffttqq] (2.9,-11.72) circle (4.5pt);
      \fill [color=ffttqq] (2.9,-10.72) circle (4.5pt);
      \fill [color=ffttqq] (-0.01,-15.76) circle (4.5pt);
      \fill [color=ffttqq] (2.9,-10.72) circle (4.5pt);
      \fill [color=ffttqq] (-2.91,-10.72) circle (4.5pt);
      \fill [color=ffttqq] (-2.91,-10.72) circle (4.5pt);
      \fill [color=ffttqq] (-0.01,-15.76) circle (4.5pt);
      \fill [color=ffttqq] (-0.01,-12.4) circle (4.5pt);
      \fill [color=ffttqq] (-0.01,-16.76) circle (4.5pt);
      \fill [color=ffttqq] (-0.01,-12.4) circle (4.5pt);
      \fill [color=ffttqq] (-3.78,-10.22) circle (4.5pt);
      \fill [color=ffttqq] (-0.01,-12.4) circle (4.5pt);
      \fill [color=ffttqq] (3.77,-10.22) circle (4.5pt);
      \fill [color=ffttqq] (-2.91,-10.72) circle (4.5pt);
      \fill [color=ffttqq] (-2.91,-10.72) circle (4.5pt);
      \fill [color=ffttqq] (-2.91,-10.72) circle (4.5pt);
      \fill [color=ffttqq] (-0.01,-15.76) circle (4.5pt);
      \fill [color=ffttqq] (-0.01,-15.76) circle (4.5pt);
      \fill [color=ffttqq] (2.9,-10.72) circle (4.5pt);
      \fill [color=ffttqq] (2.9,-10.72) circle (4.5pt);
      \fill [color=ffttqq] (10.74,5.19) circle (4.5pt);
      \fill [color=ffttqq] (10.74,6.19) circle (4.5pt);
      \fill [color=ffttqq] (11.6,6.69) circle (4.5pt);
      \fill [color=ffttqq] (10.74,6.19) circle (4.5pt);
      \fill [color=ffttqq] (9.87,6.69) circle (4.5pt);
      \fill [color=ffttqq] (10.74,6.19) circle (4.5pt);
      \fill [color=ffttqq] (9.87,6.69) circle (4.5pt);
      \fill [color=ffttqq] (11.6,6.69) circle (4.5pt);
      \fill [color=ffttqq] (10.74,5.19) circle (4.5pt);
      \fill [color=ffttqq] (11.6,6.69) circle (4.5pt);
      \fill [color=ffttqq] (10.74,5.19) circle (4.5pt);
      \fill [color=ffttqq] (9.87,6.69) circle (4.5pt);
      \fill [color=ffttqq] (10.74,2.83) circle (4.5pt);
      \fill [color=ffttqq] (10.74,1.83) circle (4.5pt);
      \fill [color=ffttqq] (10.74,2.83) circle (4.5pt);
      \fill [color=ffttqq] (11.6,3.33) circle (4.5pt);
      \fill [color=ffttqq] (10.74,2.83) circle (4.5pt);
      \fill [color=ffttqq] (11.6,3.33) circle (4.5pt);
      \fill [color=ffttqq] (10.74,1.83) circle (4.5pt);
      \fill [color=ffttqq] (10.74,1.83) circle (4.5pt);
      \fill [color=ffttqq] (9.87,3.33) circle (4.5pt);
      \fill [color=ffttqq] (11.6,3.33) circle (4.5pt);
      \fill [color=ffttqq] (13.65,6.87) circle (4.5pt);
      \fill [color=ffttqq] (13.65,7.87) circle (4.5pt);
      \fill [color=ffttqq] (14.51,8.37) circle (4.5pt);
      \fill [color=ffttqq] (13.65,7.87) circle (4.5pt);
      \fill [color=ffttqq] (12.78,8.37) circle (4.5pt);
      \fill [color=ffttqq] (13.65,7.87) circle (4.5pt);
      \fill [color=ffttqq] (12.78,8.37) circle (4.5pt);
      \fill [color=ffttqq] (14.51,8.37) circle (4.5pt);
      \fill [color=ffttqq] (13.65,6.87) circle (4.5pt);
      \fill [color=ffttqq] (14.51,8.37) circle (4.5pt);
      \fill [color=ffttqq] (13.65,6.87) circle (4.5pt);
      \fill [color=ffttqq] (12.78,8.37) circle (4.5pt);
      \fill [color=ffttqq] (8.69,8.37) circle (4.5pt);
      \fill [color=ffttqq] (7.83,7.87) circle (4.5pt);
      \fill [color=ffttqq] (6.96,8.37) circle (4.5pt);
      \fill [color=ffttqq] (7.83,7.87) circle (4.5pt);
      \fill [color=ffttqq] (7.83,6.87) circle (4.5pt);
      \fill [color=ffttqq] (7.83,7.87) circle (4.5pt);
      \fill [color=ffttqq] (7.83,6.87) circle (4.5pt);
      \fill [color=ffttqq] (6.96,8.37) circle (4.5pt);
      \fill [color=ffttqq] (8.69,8.37) circle (4.5pt);
      \fill [color=ffttqq] (6.96,8.37) circle (4.5pt);
      \fill [color=ffttqq] (8.69,8.37) circle (4.5pt);
      \fill [color=ffttqq] (7.83,6.87) circle (4.5pt);
      \fill [color=ffttqq] (10.74,6.19) circle (4.5pt);
      \fill [color=ffttqq] (12.78,8.37) circle (4.5pt);
      \fill [color=ffttqq] (10.74,6.19) circle (4.5pt);
      \fill [color=ffttqq] (13.65,6.87) circle (4.5pt);
      \fill [color=ffttqq] (10.74,6.19) circle (4.5pt);
      \fill [color=ffttqq] (11.6,3.33) circle (4.5pt);
      \fill [color=ffttqq] (10.74,6.19) circle (4.5pt);
      \fill [color=ffttqq] (9.87,3.33) circle (4.5pt);
      \fill [color=ffttqq] (10.74,6.19) circle (4.5pt);
      \fill [color=ffttqq] (7.83,6.87) circle (4.5pt);
      \fill [color=ffttqq] (10.74,6.19) circle (4.5pt);
      \fill [color=ffttqq] (8.69,8.37) circle (4.5pt);
      \fill [color=ffttqq] (7.83,7.87) circle (4.5pt);
      \fill [color=ffttqq] (13.65,7.87) circle (4.5pt);
      \fill [color=ffttqq] (7.83,7.87) circle (4.5pt);
      \fill [color=ffttqq] (10.74,2.83) circle (4.5pt);
      \fill [color=ffttqq] (10.74,2.83) circle (4.5pt);
      \fill [color=ffttqq] (13.65,7.87) circle (4.5pt);
      \fill [color=ffttqq] (10.74,6.19) circle (4.5pt);
      \fill [color=ffttqq] (14.51,8.37) circle (4.5pt);
      \fill [color=ffttqq] (10.74,6.19) circle (4.5pt);
      \fill [color=ffttqq] (10.74,1.83) circle (4.5pt);
      \fill [color=ffttqq] (10.74,6.19) circle (4.5pt);
      \fill [color=ffttqq] (6.96,8.37) circle (4.5pt);
      \fill [color=ffttqq] (10.74,6.19) circle (4.5pt);
      \fill [color=ffttqq] (10.74,6.19) circle (4.5pt);
      \fill [color=ffttqq] (10.74,6.19) circle (4.5pt);
      \fill [color=ffttqq] (10.74,6.19) circle (4.5pt);
      \fill [color=ffttqq] (10.74,6.19) circle (4.5pt);
      \fill [color=ffttqq] (10.74,6.19) circle (4.5pt);
      \fill [color=ffttqq] (10.74,6.19) circle (4.5pt);
      \fill [color=ffttqq] (10.74,6.19) circle (4.5pt);
      \fill [color=ffttqq] (10.74,6.19) circle (4.5pt);
      \fill [color=ffttqq] (10.74,6.19) circle (4.5pt);
      \fill [color=ffttqq] (10.74,6.19) circle (4.5pt);
      \fill [color=ffttqq] (10.74,6.19) circle (4.5pt);
      \fill [color=ffttqq] (10.74,2.83) circle (4.5pt);
      \fill [color=ffttqq] (10.74,2.83) circle (4.5pt);
      \fill [color=ffttqq] (10.74,2.83) circle (4.5pt);
      \fill [color=ffttqq] (13.65,7.87) circle (4.5pt);
      \fill [color=ffttqq] (13.65,7.87) circle (4.5pt);
      \fill [color=ffttqq] (7.83,7.87) circle (4.5pt);
      \fill [color=ffttqq] (7.83,7.87) circle (4.5pt);
      \fill [color=ffttqq] (-9.87,6.7) circle (4.5pt);
      \fill [color=ffttqq] (-10.73,6.2) circle (4.5pt);
      \fill [color=ffttqq] (-11.6,6.7) circle (4.5pt);
      \fill [color=ffttqq] (-10.73,6.2) circle (4.5pt);
      \fill [color=ffttqq] (-10.73,5.2) circle (4.5pt);
      \fill [color=ffttqq] (-10.73,6.2) circle (4.5pt);
      \fill [color=ffttqq] (-10.73,5.2) circle (4.5pt);
      \fill [color=ffttqq] (-11.6,6.7) circle (4.5pt);
      \fill [color=ffttqq] (-9.87,6.7) circle (4.5pt);
      \fill [color=ffttqq] (-11.6,6.7) circle (4.5pt);
      \fill [color=ffttqq] (-9.87,6.7) circle (4.5pt);
      \fill [color=ffttqq] (-10.73,5.2) circle (4.5pt);
      \fill [color=ffttqq] (-7.82,7.88) circle (4.5pt);
      \fill [color=ffttqq] (-6.96,8.38) circle (4.5pt);
      \fill [color=ffttqq] (-7.82,7.88) circle (4.5pt);
      \fill [color=ffttqq] (-8.69,8.38) circle (4.5pt);
      \fill [color=ffttqq] (-7.82,7.88) circle (4.5pt);
      \fill [color=ffttqq] (-8.69,8.38) circle (4.5pt);
      \fill [color=ffttqq] (-6.96,8.38) circle (4.5pt);
      \fill [color=ffttqq] (-6.96,8.38) circle (4.5pt);
      \fill [color=ffttqq] (-7.82,6.88) circle (4.5pt);
      \fill [color=ffttqq] (-8.69,8.38) circle (4.5pt);
      \fill [color=ffttqq] (-12.78,8.38) circle (4.5pt);
      \fill [color=ffttqq] (-13.64,7.88) circle (4.5pt);
      \fill [color=ffttqq] (-14.51,8.38) circle (4.5pt);
      \fill [color=ffttqq] (-13.64,7.88) circle (4.5pt);
      \fill [color=ffttqq] (-13.64,6.88) circle (4.5pt);
      \fill [color=ffttqq] (-13.64,7.88) circle (4.5pt);
      \fill [color=ffttqq] (-13.64,6.88) circle (4.5pt);
      \fill [color=ffttqq] (-14.51,8.38) circle (4.5pt);
      \fill [color=ffttqq] (-12.78,8.38) circle (4.5pt);
      \fill [color=ffttqq] (-14.51,8.38) circle (4.5pt);
      \fill [color=ffttqq] (-12.78,8.38) circle (4.5pt);
      \fill [color=ffttqq] (-13.64,6.88) circle (4.5pt);
      \fill [color=ffttqq] (-11.6,3.34) circle (4.5pt);
      \fill [color=ffttqq] (-10.73,2.84) circle (4.5pt);
      \fill [color=ffttqq] (-10.73,1.84) circle (4.5pt);
      \fill [color=ffttqq] (-10.73,2.84) circle (4.5pt);
      \fill [color=ffttqq] (-9.87,3.34) circle (4.5pt);
      \fill [color=ffttqq] (-10.73,2.84) circle (4.5pt);
      \fill [color=ffttqq] (-9.87,3.34) circle (4.5pt);
      \fill [color=ffttqq] (-10.73,1.84) circle (4.5pt);
      \fill [color=ffttqq] (-11.6,3.34) circle (4.5pt);
      \fill [color=ffttqq] (-10.73,1.84) circle (4.5pt);
      \fill [color=ffttqq] (-11.6,3.34) circle (4.5pt);
      \fill [color=ffttqq] (-9.87,3.34) circle (4.5pt);
      \fill [color=ffttqq] (-10.73,6.2) circle (4.5pt);
      \fill [color=ffttqq] (-13.64,6.88) circle (4.5pt);
      \fill [color=ffttqq] (-10.73,6.2) circle (4.5pt);
      \fill [color=ffttqq] (-12.78,8.38) circle (4.5pt);
      \fill [color=ffttqq] (-10.73,6.2) circle (4.5pt);
      \fill [color=ffttqq] (-8.69,8.38) circle (4.5pt);
      \fill [color=ffttqq] (-10.73,6.2) circle (4.5pt);
      \fill [color=ffttqq] (-7.82,6.88) circle (4.5pt);
      \fill [color=ffttqq] (-10.73,6.2) circle (4.5pt);
      \fill [color=ffttqq] (-9.87,3.34) circle (4.5pt);
      \fill [color=ffttqq] (-10.73,6.2) circle (4.5pt);
      \fill [color=ffttqq] (-11.6,3.34) circle (4.5pt);
      \fill [color=ffttqq] (-10.73,2.84) circle (4.5pt);
      \fill [color=ffttqq] (-13.64,7.88) circle (4.5pt);
      \fill [color=ffttqq] (-10.73,2.84) circle (4.5pt);
      \fill [color=ffttqq] (-7.82,7.88) circle (4.5pt);
      \fill [color=ffttqq] (-7.82,7.88) circle (4.5pt);
      \fill [color=ffttqq] (-13.64,7.88) circle (4.5pt);
      \fill [color=ffttqq] (-10.73,6.2) circle (4.5pt);
      \fill [color=ffttqq] (-14.51,8.38) circle (4.5pt);
      \fill [color=ffttqq] (-10.73,6.2) circle (4.5pt);
      \fill [color=ffttqq] (-6.96,8.38) circle (4.5pt);
      \fill [color=ffttqq] (-10.73,6.2) circle (4.5pt);
      \fill [color=ffttqq] (-10.73,1.84) circle (4.5pt);
      \fill [color=ffttqq] (-10.73,6.2) circle (4.5pt);
      \fill [color=ffttqq] (-10.73,6.2) circle (4.5pt);
      \fill [color=ffttqq] (-10.73,6.2) circle (4.5pt);
      \fill [color=ffttqq] (-10.73,6.2) circle (4.5pt);
      \fill [color=ffttqq] (-10.73,6.2) circle (4.5pt);
      \fill [color=ffttqq] (-10.73,6.2) circle (4.5pt);
      \fill [color=ffttqq] (-10.73,6.2) circle (4.5pt);
      \fill [color=ffttqq] (-10.73,6.2) circle (4.5pt);
      \fill [color=ffttqq] (-10.73,6.2) circle (4.5pt);
      \fill [color=ffttqq] (-10.73,6.2) circle (4.5pt);
      \fill [color=ffttqq] (-10.73,6.2) circle (4.5pt);
      \fill [color=ffttqq] (-10.73,6.2) circle (4.5pt);
      \fill [color=ffttqq] (-7.82,7.88) circle (4.5pt);
      \fill [color=ffttqq] (-7.82,7.88) circle (4.5pt);
      \fill [color=ffttqq] (-7.82,7.88) circle (4.5pt);
      \fill [color=ffttqq] (-13.64,7.88) circle (4.5pt);
      \fill [color=ffttqq] (-13.64,7.88) circle (4.5pt);
      \fill [color=ffttqq] (-10.73,2.84) circle (4.5pt);
      \fill [color=ffttqq] (-10.73,2.84) circle (4.5pt);
    \end{scriptsize}
  \end{tikzpicture}
  \label{fig:hierarchical}
  }
  \qquad
  \subfigure[]{
    \includegraphics[width=5cm]{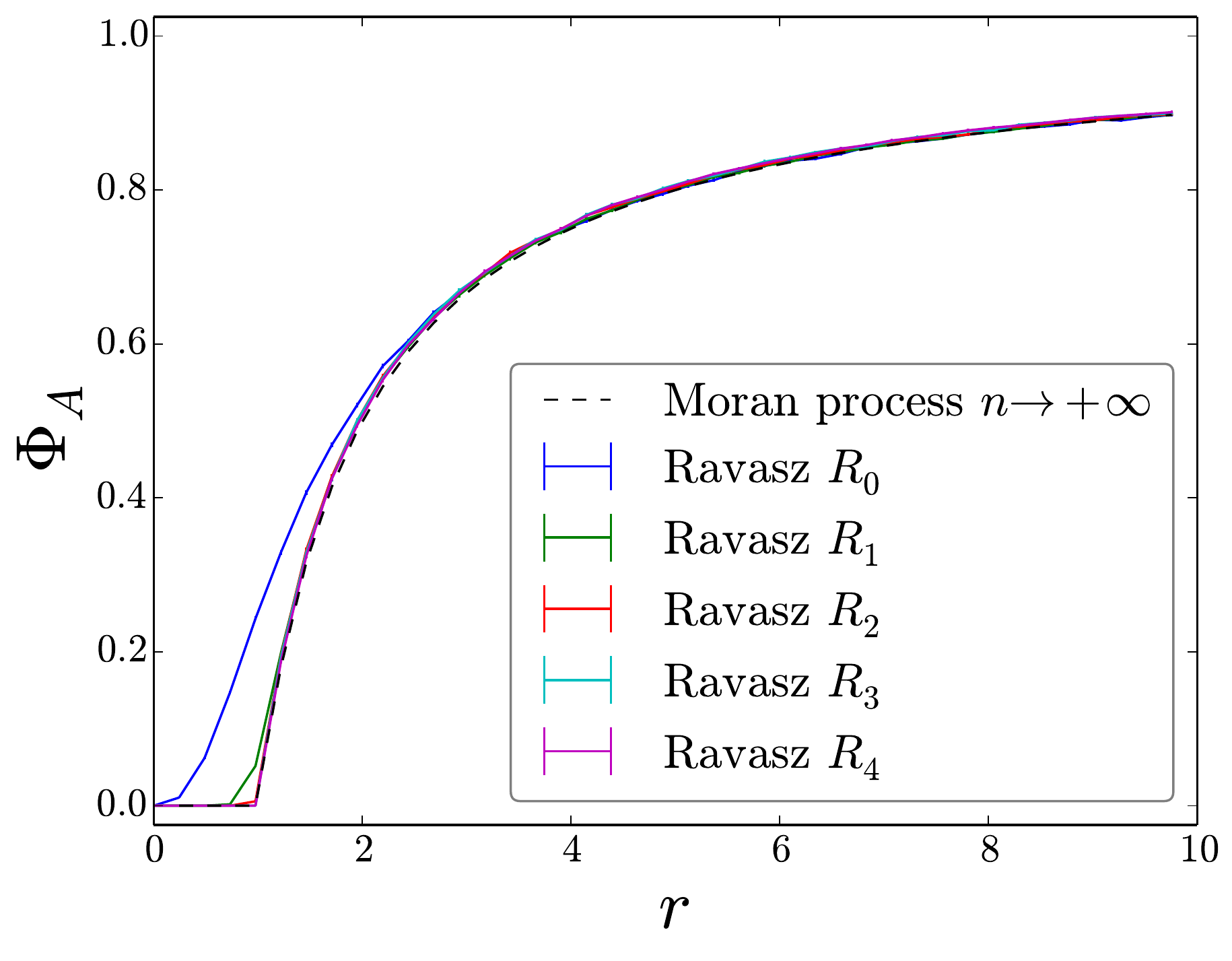}
    \label{fig:hierarchical_fix}
  }
  \caption{(a) Third step in the construction of a 
    hierarchical network by~\cite{RSMOB}, and (b) fixation probability for
    different construction steps compared with the Moran process as 
    $n\to+\infty$. Here $100\,000$ trials was carried out to obtain a less 
    noisy approximation of the fixation probability.}
\end{figure}

In~\cite{RSMOB}, a deterministic network was introduced as a heuristic model of metabolic networks (it can be seen in Figure~\ref{fig:hierarchical}). The 
graph has a power law distribution of the degree (scale-free topology) and mean 
clustering coefficient non-decreasing with size.

The network is constructed inductively. In the first step, we define the 
network $R_0$ as the complete graph of order four $K_4$. Fix one vertex as the 
\emph{central vertex}. The rest of the vertices are \emph{external vertices}. 
Now, take three copies of $R_0$ and join their external vertices with the 
central one of the original $R_0$ and their central vertices together making 
a big triangle. The central vertex of the resulting graph $R_1$ is the central 
vertex of the original $R_0$. The external vertices of $R_1$ are the vertices 
of the other copies of $R_0$. You can repeat the process as many times as 
needed, obtaining graphs $R_n$ of order $4^{n+1}$.

We computed the average fixation probability and the average computation time
on this family for the same 
values of fitness and trials as before. The results can be seen in Figure~\ref{fig:hierarchical_fix} and Table~\ref{tab:ravasz}. As one can see, the EMC method outperforms the SMC method by an increasing factor on the order $n$ of the network. 
%
\begin{table}
  \caption{Average computation times (in seconds) for Moran processes on the 
    hierarchical model by \cite{RSMOB} using EMC and SMC 
    methods, both with $1000$ trials and 
    $r$ going from $0$ to $10$ with step size of $0.25$. The last 
    column shows how many times faster EMC is than the SMC method.}
  \label{tab:ravasz}
  \begin{center}
    \begin{tabular}{ccccc}
      \toprule
      Step &  Graph order  & EMC time &  SMC time & SMC/EMC \\
      \midrule
       $0$ & $\invi{121}4$ & $\invi{70}0.76$ & $\invi{692}0.82$ & $1.07$\\ 
       $1$ & $\invi{12}16$ & $\invi{70}0.88$ & $\invi{692}0.94$ & $1.06$\\ 
       $2$ & $\invi{12}64$ & $\invi{70}1.92$ & $\invi{692}5.15$ & $2.68$\\ 
       $3$ & $\invi{1}256$ & $\invi{7}24.94$ & $\invi{6}141.24$ & $5.66$\\ 
       $4$ & $       1024$ & $       707.61$ & $       6923.04$ & $9.78$\\
      \bottomrule
    \end{tabular}
  \end{center}
\end{table}

\section{Conclusion}

In this paper, we review some fundamental ideas and results on evolutionary 
dynamics as introduced in~\cite{LHN} generalising the classic process 
described in~\cite{M}. But we also give insights on one of the major problems 
in this theory, to estimate the average fixation probability of a mutant with 
relative fitness $r$ on a given graph. Exact solutions have only been computed 
for a few families of graphs~\cite{BR} as, generally, one should solve a 
linear equation systems of $2^N$ equations, where $N$ is the order of the 
graph. Even asymptotic behaviour is tricky to compute. 

The erasing of loops in the state space is the geometrical counterpart of a 
well known device in Markov chains, which is the basis behind embedded 
processes and which consists of forcing the processes to evolve in each 
iteration. It is rather obvious and well known that the expected fixation time 
is reduced and the fixation probability is unchanged by this procedure. 
In this paper,  we use this idea to compute asymptotically the fixation 
probability for the class of complete bipartite graphs, generalising the result 
of~\cite{LHN} for the star graph. In this case, the high degree symmetry 
reduces the problem to a set of 
$2N$ equations, which is asymptotically equivalent to a simpler linear system 
of $N$ equations. For complete bipartite graphs, after erasing all non-trivial 
loops, partial symmetries arise asymptotically in the Moran process 
and reduce the Moran process to the particular case of a star graph. 
This is an important step since it shed some new light on the asymptotic 
behaviour of the fixation on bipartite graphs, which has recently been dealt with from other points of view in~\cite{Vallade} and~\cite{TanLu}.

In practice, the Monte Carlo on the Embedded Markov chain (EMC method) may need 
to make more computations than the Monte 
Carlo method on the standard chain (SMC method), as it needs to keep track of 
different probabilities (one per vertex) that should be computed at runtime. We 
tested the speed of the new method in some celebrated families of graphs: the 
small world networks~\cite{WS}, preferential attachment networks~\cite{BA} 
and hierarchical networks~\cite{RSMOB}. These tests show the EMC method 
defeats the SMC method on large families of graphs, but not in all examples, as 
transition probabilities on the loop-erased chain depends heavily on the actual 
state. At first, the appearance of high degree vertices might look like culprit 
for this problem, but this is not the case: the hierarchical 
network of~\cite{RSMOB} has extremely large degree on some vertices. Although it is still 
unknown what makes EMC method become slower, we believe this method could be 
applied successfully to real networks. 

Globally, we think the present paper represents substantial progress towards 
understanding the complexity behind evolutionary dynamics on graphs.

\section*{Acknowledgment}

This research was partially supported by the Ministry of Science and Innovation 
- Government of Spain (Grant MTM2010-15471) and IEMath Network CN 2012/077. 
Last author was also supported by the European Social Fund and Diputaci\'on General de Arag\'on (Grant E15 Geometr\'ia).

The authors thank two anonymous reviewers for their accurate comments.


\begin{thebibliography}{10}

\bibitem{Banerjee}
A.~Banerjee.
\newblock Structural distance and evolutionary relationship of networks.
\newblock {\em Biosystems}, 107(3):186 -- 196, 2012.

\bibitem{BA}
A.-L. Barab{\'a}si and R.~Albert.
\newblock Emergence of scaling in random networks.
\newblock {\em Science}, 286:509--512, 1999.

\bibitem{BR}
M.~Broom and J.~Rycht{{\'a}}{\v{r}}.
\newblock An analysis of the fixation probability of a mutant on special
  classes of non-directed graphs.
\newblock {\em Proc. R. Soc. Lond. Ser. A Math. Phys. Eng. Sci.},
  464(2098):2609--2627, 2008.

\bibitem{BRS1}
M.~Broom, J.~Rycht{{\'a}}{\v{r}}, and B.~T. Stadler.
\newblock Evolutionary dynamics on small-order graphs.
\newblock {\em J. Intesdiscip. Math.}, 12:129--140, 2009.

\bibitem{BRS2}
M.~Broom, J.~Rycht{{\'a}}{\v{r}}, and B.~T. Stadler.
\newblock Evolutionary dynamics on graphs---the effect of graph structure and
  initial placement on mutant spread.
\newblock {\em J. Stat. Theory Pract.}, 5(3):369--381, 2011.

\bibitem{D&al}
J.~D\'{\i}az, L.~A. Goldberg, G.~B. Mertzios, D.~Richerby, M.~Serna, and P.~G.
  Spirakis.
\newblock Approximating fixation probabilities in the generalized moran
  process.
\newblock In {\em Proceedings of the Twenty-Third Annual ACM-SIAM Symposium on
  Discrete Algorithms}, SODA '12, pages 954--960. SIAM, 2012.

\bibitem{H}
C.~C. Hadjichrysanthou.
\newblock {\em Evolutionary models in structured populations}.
\newblock PhD thesis, City University London, 2012.

\bibitem{HSS}
A.~A. Hagberg, D.~A. Schult, and P.~J. Swart.
\newblock Exploring network structure, dynamics, and function using {NetworkX}.
\newblock In {\em Proceedings of the 7th Python in Science Conference
  (SciPy2008)}, pages 11--15, Pasadena, CA USA, Aug. 2008.

\bibitem{Vallade}
B.~Houchmandzadeh and M.~Vallade.
\newblock Exact results for fixation probability of bithermal evolutionary
  graphs.
\newblock {\em Biosystems}, 112(1):49 -- 54, 2013.

\bibitem{KT}
S.~Karlin and H.~M. Taylor.
\newblock {\em A First Course in Stochastic Processes}.
\newblock Academic Press Inc., New York, N.Y., second edition, 1975.

\bibitem{LHN}
E.~Lieberman, C.~Hauert, and M.~A. Nowak.
\newblock Evolutionary dynamics on graphs.
\newblock {\em Nature}, 433(7023):312--316, Jan. 2005.

\bibitem{M}
P.~A.~P. Moran.
\newblock Random processes in genetics.
\newblock {\em Proc. Cambridge Philos. Soc.}, 54:60--71, 1958.

\bibitem{N}
M.~A. Nowak.
\newblock {\em Evolutionary Dynamics: Exploring the Equations of Life}.
\newblock Belknap Press of Harvard University Press, Sept. 2006.

\bibitem{Nowak&al}
M.~A. Nowak, A.~Sasaki, C.~Taylor, and D.~Fudenberg.
\newblock Emergence of cooperation and evolutionary stability in finite
  populations.
\newblock {\em Nature}, 428(6983):646--650, 2004.

\bibitem{RSMOB}
E.~Ravasz, A.~L. Somera, D.~A. Mongru, Z.~N. Oltvai, and A.-L. Barab{\'a}si.
\newblock Hierarchical organization of modularity in metabolic networks.
\newblock {\em Science (New York, N.Y.)}, 297(5586):1551--1555, Aug. 2002.

\bibitem{RS}
J.~Rycht{{\'a}}{\v{r}} and B.~Stadler.
\newblock Evolutionary dynamics on small-world networks.
\newblock {\em International Journal of Computational and Mathematical Sciences
  [electronic only]}, 2(1):1--4, electronic only, 2008.

\bibitem{SR1}
P.~Shakarian and P.~Roos.
\newblock Fast and deterministic computation of fixation probability in
  evolutionary graphs.
\newblock In {\em In: CIB '11: The Sixth IASTED Conference on Computational
  Intelligence and Bioinformatics (accepted). IASTED}, 2011.

\bibitem{SR2}
P.~Shakarian, P.~Roos, and A.~Johnson.
\newblock A review of evolutionary graph theory with applications to game
  theory.
\newblock {\em Biosystems}, 107(2):66 -- 80, 2012.

\bibitem{ShakarianBio}
P.~Shakarian, P.~Roos, and G.~Moores.
\newblock A novel analytical method for evolutionary graph theory problems.
\newblock {\em Biosystems}, 111(2):136 -- 144, 2013.

\bibitem{sage}
W.~A. Stein et~al.
\newblock {\em {S}age {M}athematics {S}oftware ({V}ersion 5.8)}.
\newblock The Sage Development Team, 2013.
\newblock {\tt http://www.sagemath.org}.

\bibitem{TanLu}
S.~Tan and J.~Lu.
\newblock Characterizing the effect of population heterogeneity on evolutionary
  dynamics on complex networks.
\newblock {\em Sci. Rep.}, 4, may 2014.

\bibitem{Taylor&al}
C.~Taylor, D.~Fudenberg, A.~Sasaki, and M.~Nowak.
\newblock Evolutionary game dynamics in finite populations.
\newblock {\em Bulletin of Mathematical Biology}, 66(6):1621--1644, 2004.

\bibitem{KT:ISM}
H.~M. Taylor and S.~Karlin.
\newblock {\em An introduction to stochastic modeling}.
\newblock Academic Press Inc., San Diego, CA, third edition, 1998.

\bibitem{WS}
D.~J. Watts and S.~H. Strogatz.
\newblock Collective dynamics of `small-world' networks.
\newblock {\em Nature}, 393(6684):440--442, June 1998.

\end{thebibliography}
\end{document}